\documentclass[pdflatex,sn-mathphys-num,iicol]{sn-jnl}

\usepackage{graphicx}%
\usepackage{multirow}%
\usepackage{amsmath,amssymb,amsfonts}%
\usepackage{amsthm}%
\usepackage{mathrsfs}%
\usepackage[title]{appendix}%
\usepackage{xcolor}%
\usepackage{textcomp}%
\usepackage{manyfoot}%
\usepackage{booktabs}%
\usepackage{algorithm}%
\usepackage{algorithmicx}%
\usepackage{algpseudocode}%
\usepackage{listings}%

\usepackage[makeroom]{cancel}
\usepackage{scalerel,stackengine}

\newcommand{\transpose}{{\mathsf{T}}}
\newcommand{\lagrange}{\mathord{\mathrel{\raisebox{-2.25pt}{$\mathcal{P}$}}}}

\newcommand{\frobenius}{\mathrel{\text{\raisebox{1.0pt}{$:$}}}}

\theoremstyle{thmstyleone}%
\theoremstyle{thmstyletwo}%

\theoremstyle{thmstylethree}%

\begin{document}

\title[Modelling Material Injection]{Modelling Material Injection Into Porous Structures Under Non-isothermal Conditions}


\author*[1,2]{\fnm{J.-S. L.} \sur{V\"olter}}\email{jan-soeren.voelter@imsb.uni-stuttgart.de}

\author[2]{\fnm{Z.} \sur{Trivedi}}

\author[3]{\fnm{A.} \sur{Boger}}

\author[2]{\fnm{T.} \sur{Ricken}}

\author[1,4]{\fnm{O.} \sur{R\"ohrle}}

\affil[1]{\orgdiv{Institute for Modelling and Simulation of Biomechanical Systems}, \orgname{University of Stuttgart}, \orgaddress{\street{Pfaffenwaldring 5a}, \city{Stuttgart}, \postcode{70569}, \country{Germany}}}

\affil[2]{\orgdiv{Institute of Structural Mechanics and Dynamics of Aerospace Structures}, \orgname{University of Stuttgart}, \orgaddress{\street{Pfaffenwaldring 27}, \city{Stuttgart}, \postcode{70569}, \country{Germany}}}

\affil[3]{\orgdiv{Biomechanical Engineering}, \orgname{Ansbach University of Applied Sciences}, \orgaddress{\street{Residenzstra{\ss}e 8}, \city{Ansbach}, \postcode{91522}, \country{Germany}}}

\affil[4]{\orgdiv{Stuttgart Center for Simulation Science (SC SimTech)}, \orgname{University of Stuttgart}, \orgaddress{\street{Pfaffenwaldring 5a}, \city{Stuttgart}, \postcode{70569}, \country{Germany}}}


\abstract{In this work, the Theory of Porous Media (TPM) is employed to model percutaneous vertebroplasty, a medical procedure in which acrylic cement is injected into cancellous vertebral bone. Previously, isothermal macroscale models have been derived to describe this material injection and the arising mechanical interactions. However, the temperature of the injected cement is typically below the human body temperature, necessitating the extension of these existing models to the non-isothermal case. Following the modelling principles of the TPM and consi\-dering local thermal non-equilibrium conditions, our model introduces three energy balances as well as constitutive relations for thermal conduction and heat transfer. If restricted to local thermal equilibrium conditions, our model equations are in agreement with other TPM-based models. We observe that our model elicits physically reasonable behaviour in numerical simulations that employ parameter values and initial and boundary conditions relevant for our application. Noting that we neglect capillary effects, we claim our model to be thermodynamically consistent despite the employment of simplifying assumptions during its derivation, such as the Coleman and Noll procedure.}




\maketitle

\section{Introduction}\label{sec1}

Percutaneous vertebroplasty (PV) is a medical procedure in which bio-compatible cement is injected into structurally unstable or damaged vertebrae~\cite{jensen1997}. Thereafter, the cement undergoes curing, stabilising the vertebra mechanically. Many complications during PV are associated with the leakage of cement into extra-vertebral tissue~\cite{laredo2004}. The consequences can be fatal, such that it is desirable to predict the outcome of PV pre-operatively.\\
To this extent, continuum-mechanical models have been developed to describe and simulate the injection of cement. Pore-scale simulations have been performed previously, both with and without consi\-deration of the cement curing~\cite{Zeiser08,Kolmeder2012,Landgraf15,Kolmeder16}. Inherently, these microscale models are computationally expensive, limiting the spatial extent of the considered geometries. Macroscale models that we presented previously~\cite{Bleiler15,Trivedi2022}, have been based on the Theory of Porous Media (TPM). Inherently, these models describe the injection of cement into a vertebral body as a fully coupled mechanical problem~\cite{Bleiler15}. However, they assume isothermal conditions. This contradicts the boundary conditions of the cement injection during PV. In particular, the initial temperature of the cement differs significantly from the human body temperature. Further, the cement curing process is exothermic, generating heat.\\
In this work, we extend these TPM-based models to the non-isothermal case, considering local thermal non-equilibrium (LTNE) conditions as opposed to local thermal equilibrium (LTE) conditions. We only consider preliminary assumptions and material parameter values relevant for the modelling of PV.\\
Non-isothermal formulations based on the TPM can also be taken from the works \cite{Ricken.2010b, Bluhm.2014}, focusing on other application areas.

\section{Methods}\label{secMethods}
\subsection{Fundamentals of the TPM}
In the following, we provide the necessary fundamentals of the TPM. For a comprehensive overview see, e.g., the works of Ehlers~\cite{Ehlers02,ehlers2009}.
\subsubsection{Multiphasic Description}
\begin{figure}[htbp]
\includegraphics[scale=0.128]{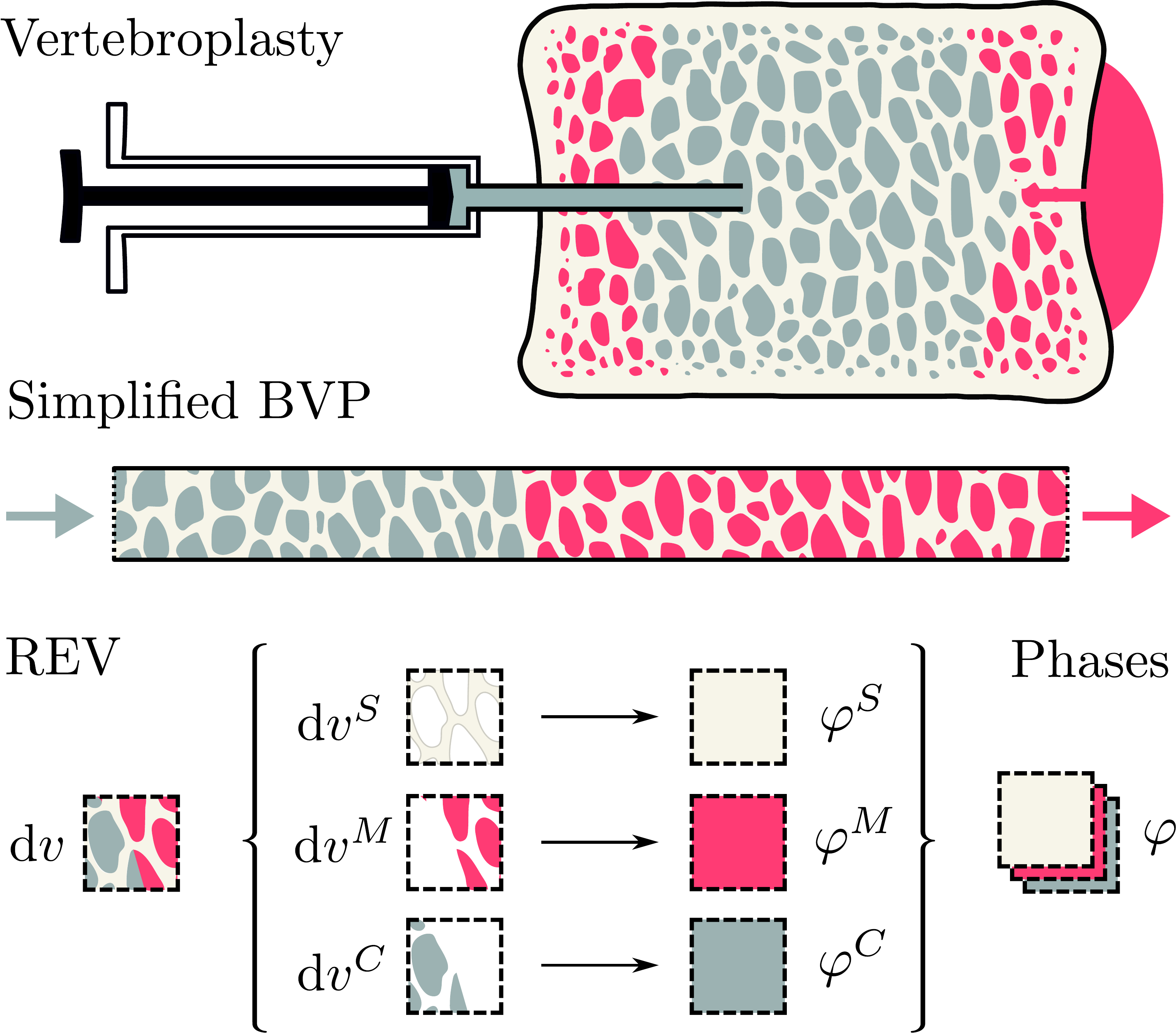}
\caption{Injection of cement into a vertebral body, simplified boundary value problem (BVP) as well as an overview of the three constituents which are considered. The pore size is greatly exaggerated for readability}
\label{fig:overview}
\end{figure}

The TPM describes porous media based on representative elementary volumes (REVs) whose microstructure is homogenised in the sense of volumetric averaging. This allows a macroscopic description of porous media, where constituents are treated as spatially superimposed continua referred to as phases. Here, three constituents are considered, yielding a solid phase $\varphi^S$ and two immiscible fluid phases $\varphi^M$ and $\varphi^C$. They represent trabecular bone, bone marrow and bone cement, respectively, as visualised in Figure~\ref{fig:overview}.

The local composition of the porous medium is captured by volume fractions 
\begin{equation}
n^{\alpha}: = \frac{\mathrm{d}v^{\alpha}}{\mathrm{d}v},\quad \alpha\in\{S,M,C\},
\end{equation}
defined by the ratio of the partial constituent volumes $\mathrm{d}v^\alpha$ and the aggregate volume $\mathrm{d}v$. Under fully saturated conditions, it follows that
\begin{equation}\label{eq:saturation}
\sum_\alpha n^\alpha = 1.
\end{equation}
Further, fluid saturations are defined as
\begin{equation}
s^\beta := \frac{n^\beta}{n^F} ,\quad \beta\in\{M,C\},
\end{equation}
where
\begin{equation}
n^F := \sum_\beta n^\beta,\quad\mathrm{such\ that}\quad \sum_\beta s^\beta = 1.
\end{equation}
Denoting their local masses as $\mathrm{d}m^\alpha$ each constituent is assigned a material and an apparent density defined as
\begin{equation}
\rho^{\alpha\mathrm{R}} := \frac{\mathrm{d}m^\alpha}{\mathrm{d}v^\alpha}\quad\mathrm{and}\quad \rho^\alpha:=\frac{\mathrm{d}m^\alpha}{\mathrm{d}v},
\end{equation}
respectively.
The description of motion of porous media borrows from the description of single-phase continua, i.e., the current configuration of a continuum body is uniquely determined by defining a reference configuration and evolving it in time according to the continuum's motion function. The position $\mathbf{X}_{\alpha}$ of material points of constituent $\varphi^\alpha$ evolves according to the unique motion path $\boldsymbol{\chi}_{\alpha}$. Therewith, the position of spatial points at time $t$ is defined as
\begin{equation}
\mathbf{x}(t) := \boldsymbol{\chi}_{\alpha}(\mathbf{X}_{\alpha},t).
\end{equation}
Since all phases occupy the same space, each spatial point $\mathbf{x}$ can be defined with respect to any constituent. Considering derivatives at spatial points, constituent-specific velocities are defined as
\begin{equation}
\overset{\prime}{\mathbf{x}}_{\alpha} := (\mathbf{x})^\prime_\alpha := \frac{\mathrm{d} \boldsymbol{\chi}_{\alpha}(\mathbf{X}_{\alpha},t)}{\mathrm{d} t}.
\end{equation}
In our model, solid motion is expressed via the solid displacement
\begin{equation}
\mathbf{u}_{S} := \mathbf{x}-\mathbf{X}_{S},\quad\mathrm{such\ that}\quad(\mathbf{u}_S)^\prime_S = \overset{\prime}{\mathbf{x}}_{S},
\end{equation}
and fluid motion via the seepage velocities
\begin{equation}
\mathbf{w}_{\beta} := \overset{\prime}{\mathbf{x}}_{\beta} -\overset{\prime}{\mathbf{x}}_{S}.
\end{equation}
Constituent-specific material deformation gradients are defined by
\begin{equation}
\mathbf{F}_{\alpha} := \frac{\partial \boldsymbol{\chi}_{\alpha}(\mathbf{X}_{\alpha},t)}{\partial \mathbf{X}_{\alpha}},
\end{equation}
with which further constituent-specific deformation and strain measures can be defined, e.g. the right Cauchy-Green tensor $\mathbf{C}_S := \mathbf{F}_S^\transpose \mathbf{F}_S$. 

\subsubsection{Balance Laws}
Following Truesdell's metaphysics (cf.~\cite{Ehlers02}), the aggregate is treated as a continuum body, i.e., it is axiomatically assigned balance laws governing the conservation of mass, linear momentum, angular momentum, energy and the non-negativity of total entropy production. Analogously, each constituent is axiomatically assigned a constituent mass balance
\begin{equation}\label{eq:constituentmass}
(\rho^\alpha)^\prime_\alpha + \rho^\alpha \mathrm{div}\,\overset{\prime}{\mathbf{x}}_\alpha = \hat{\rho}^\alpha,
\end{equation}
a constituent linear momentum balance
\begin{equation}
\rho^\alpha \overset{\prime\prime}{\mathbf{x}}_\alpha = \mathrm{div}\,\mathbf{T}^\alpha + \rho^\alpha \mathbf{b}^\alpha + \hat{\mathbf{p}}^\alpha,
\end{equation}
a constituent angular momentum balance
\begin{equation}
\mathbf{0} = \mathbf{I}\times\mathbf{T}^\alpha + \hat{\mathbf{m}}^\alpha,
\end{equation}
a constituent energy balance
\begin{equation}\label{eq:energy}
\rho^\alpha (\varepsilon^\alpha)^\prime_\alpha = \mathbf{T}^\alpha \frobenius \mathbf{L}_\alpha - \mathrm{div}\,\mathbf{q}^\alpha + \rho^\alpha r^\alpha + \hat{\varepsilon}^\alpha
\end{equation}
and a constituent entropy balance
\begin{equation}
\rho^\alpha (\eta^\alpha)^\prime_\alpha = -\mathrm{div}(\frac{1}{\theta^\alpha}\mathbf{q}^\alpha) + \frac{1}{\theta^\alpha} \rho^\alpha r^\alpha + \hat{\zeta}^\alpha.
\end{equation}
Therein, $\mathbf{T}^\alpha$ denotes the partial Cauchy stress, $\mathbf{b}^\alpha$ a body force, $\mathbf{L}_\alpha$ the spatial velocity gradient, $\varepsilon^\alpha$ the internal energy, $\mathbf{q}^\alpha$ the heatflux vector, $r^\alpha$ the external heat supply, $\eta^\alpha$ the entropy and $\theta^\alpha$ the absolute temperature of constituent $\varphi^\alpha$. The hatted terms $\hat{\rho}^\alpha$, $\hat{\mathbf{p}}^\alpha$, $\hat{\mathbf{m}}^\alpha$, $\hat{\varepsilon}^\alpha$ and $\hat{\zeta}^\alpha$ denote the direct mass, linear momentum, angular momentum, energy and entropy production of constituent $\varphi^\alpha$.
The sum of the constituent balance laws is required to be identical to the corresponding aggregate balance laws. This yields restrictions for constituent specific terms and quantities. In particular, for the momentum and energy production terms one can recover
\begin{equation}\label{eq:momenumconstr}
\sum_\alpha \left(\hat{\mathbf{p}}^\alpha + \hat{\rho}^\alpha \overset{\prime}{\mathbf{x}}_\alpha\right) = \mathbf{0}\quad\text{and}\quad \sum_\alpha \hat{e}^\alpha = 0
\end{equation}
with
\begin{equation}
\hat{e}^\alpha := \hat{\varepsilon}^\alpha + \hat{\mathbf{p}}^\alpha\cdot\overset{\prime}{\mathbf{x}}_\alpha + \hat{\rho}^\alpha(\varepsilon^\alpha + {\textstyle\frac{1}{2}}\,\overset{\prime}{\mathbf{x}}_\alpha\cdot\overset{\prime}{\mathbf{x}}_\alpha).
\end{equation}

\subsubsection{Clausius-Duhem Inequality}
By combining the entropy balances with the energy balances of the porous medium and summation over all constituents, an overall entropy inequality is derived as
\begin{equation}
\begin{split}
  & \sum_{\alpha} \frac{1}{\theta^{\alpha}}\Big(\mathbf{T}^{\alpha}\frobenius\mathbf{L}_{\alpha} - \rho^{\alpha}\,(\psi^{\alpha})^{\prime}_{\alpha} - \rho^{\alpha}\,(\theta^{\alpha})^{\prime}_{\alpha}\,\eta^{\alpha}
  \\
  & -\frac{1}{\theta^{\alpha}}\,\mathbf{q}^{\alpha}\cdot\mathrm{grad}\,\theta^{\alpha}- \hat{\mathbf{p}}^{\alpha}\cdot\overset{\prime}{\mathbf{x}}_{\alpha} \hphantom{\Big) \geq 0}
  \\
  &  -\hat{\rho}^{\alpha}(\psi^{\alpha}+{\textstyle\frac{1}{2}}\,\overset{\prime}{\mathbf{x}}_{\alpha}\cdot\overset{\prime}{\mathbf{x}}_{\alpha}) + \hat{e}^{\alpha}\Big) \geq 0.
\end{split}
\end{equation}
This is the Clausius-Duhem inequality for non-polar materials. Therein, the internal energy is Legendre transformed yielding
\begin{equation}\label{equation:legendre1}
\varepsilon^{\alpha} = \psi^{\alpha} + \theta^{\alpha}\,\eta^{\alpha},
\end{equation}
introducing the constituent Helmholtz potentials or Helmholtz free energies $\psi^\alpha$.
\subsection{Modelling Assumptions}\label{sec:assumptions}
In the following, we provide the modelling assumptions leading to our model.
\subsubsection{Preliminary Assumptions}
Following the principles of the TPM~\cite{Ehlers02}, our application admits the following assumptions: the mass production $\hat{\rho}^\alpha$ is neglected, the angular momentum production $\hat{\mathbf{m}}^\alpha$ is neglected, the body forces $\mathbf{b}^\alpha$ are assumed to be identical, constant and uniform, the material densities $\rho^{\alpha\mathrm{R}}$ are assumed to be constant and uniform, and quasi-static conditions are assumed, such that the accelerations $\overset{\prime\prime}{\mathbf{x}}_{\alpha}$  are neglected. In contrast to the isothermal approaches~\cite{Bleiler15,Trivedi2022}, we assume LTNE conditions. With these modelling assumptions, the Clausius-Duhem inequality can be rewritten as
\begin{equation}\label{eq:entropy0}
\begin{split}
  & \sum_{\alpha} \frac{\theta^S}{\theta^{\alpha}}\Big(\mathbf{T}^{\alpha}\!\frobenius\mathbf{L}_{\alpha} - \rho^{\alpha}(\psi^{\alpha})^{\prime}_{\alpha} - \rho^{\alpha}(\theta^{\alpha})^{\prime}_{\alpha}\,\eta^{\alpha}\Big)
  \\
  & - \sum_{\beta} \hat{\mathbf{p}}^{\beta}\cdot\mathbf{w}_{\beta} + \sum_\beta \hat{\varepsilon}^\beta\,\frac{1}{\theta^\beta}(\theta^S-\theta^\beta)
  \\
  & - \sum_{\alpha} \frac{\theta^S}{\theta^{\alpha}}\frac{1}{\theta^\alpha}\mathbf{q}^\alpha\cdot\mathrm{grad}\,\theta^\alpha \geq 0.
\end{split}
\end{equation}
For the sake of completeness, with the above assumptions the constituent mass balances reduce to volume balances and the solid volume fraction is fully determined by
\begin{equation}
n^S = n^S_{0S} (\mathrm{det}\,\mathbf{F}_S)^{-1},
\end{equation}
where $n^S_{0S}$ denotes the value of the solid volume fraction in the reference configuration of the solid phase. Further, the balances of angular momentum reduce to symmetry requirements for the Cauchy stress tensors and are removed as independent equations in the context of the closure problem of continuum mechanics.
\subsubsection{Assumptions for the Helmholtz Potentials}
To evolve the entropy inequality~(\ref{eq:entropy0}) further, the dependence of the Helmholtz potentials on the independent variables has to be specified constitutively. In this work, we employ the principle of phase separation, which is claimed not to contradict the fundamental principles of the TPM~\cite{Ehlers02}.\\ 
In particular, each Helmholtz potential $\psi^\alpha$ shall depend on the respective constituent's temperature $\theta^\alpha$. Further, $\psi^S$ shall depend on the solid deformation gradient $\mathbf{F}_S$. Following examples of the TPM~\cite{ehlers2009}, only $\psi^M$ shall depend on the marrow saturation $s^M$, since $\varphi^M$ is the wetting fluid in our application. In agreement with examples of the TPM~\cite{Ehlers02,ehlers2009,Graf08}, we do not consider second grade materials and omit the spatial gradients of the above variables. Likewise, we do not consider the Helmholtz potentials to depend on the seepage velocities or the spatial velocity gradients.\\
Typically, at this point of the derivation, the saturation constraint~(\ref{eq:saturation}) is incorporated into the entropy inequality by means of the method of Lagrange multipliers~(e.g.~\cite{ehlers2009, Bleiler12}). Ultimately, this results in the introduction of effective constituent pressures.\\
Within a framework deviating from the TPM, the dependence of the Helmholtz potentials on the material densities is assumed in order to be able to directly introduce the constituent pressures based on the fundamental thermodynamic relation~(cf.~\cite{Hassanizadeh90}). Thermodynamic pressure is thereby defined as
\begin{equation}\label{eq:thermodynamicpressure0}
p^{\alpha\mathrm{R}} := -\frac{\partial \psi^\alpha}{\partial v^{\alpha\mathrm{R}}} = (\rho^{\alpha\mathrm{R}})^2\,\frac{\partial \psi^\alpha}{\partial \rho^{\alpha\mathrm{R}}},
\end{equation}
with $v^{\alpha\mathrm{R}}:=1/\rho^{\alpha\mathrm{R}}$. We combine both approaches. In particular, we introduce three Lagrange multipliers $\lagrange^\alpha$ and add the equality constraints
\begin{equation}\label{eq:lagrange}
0 = \frac{\theta^S}{\theta^\alpha} \lagrange^\alpha\! \left[ n^\alpha \mathbf{I}\frobenius\mathbf{L}_\alpha\! + (n^\alpha)^\prime_\alpha\! + \frac{n^\alpha}{\rho^{\alpha\mathrm{R}}} (\rho^{\alpha\mathrm{R}})^\prime_\alpha \right]
\end{equation}
to the overall entropy inequality, wherein $\alpha\in\{S,M,C\}$. These constraints represent the constituent mass balances~(\ref{eq:constituentmass}). Further, we formally consider the dependence of each Helmholtz potential on the respective specific volume $v^{\alpha\mathrm{R}}$ and note that this has no consequences for the final model since we assume material incompressibility.\\
In summary, the dependence of the Helmholtz potentials is in agreement with the assumptions seen in typical examples of TPM-based models~\cite{Ehlers02,ehlers2009,Graf08}. Exploiting algebraic relations, we rearrange the entropy inequality of our model as
\begin{subequations}
\begin{align}
& \Big(\mathbf{T}^S+n^S \lagrange^{S}\mathbf{I}-\rho^S\frac{\partial\psi^S}{\partial \mathbf{F}_S}\mathbf{F}_S^\transpose\Big)\frobenius\mathbf{L}_{S} \vphantom{\sum_\beta \frac{\theta^S}{\theta^\beta}} \label{equation:entropy-solid} \\
& + \sum_\beta \frac{\theta^S}{\theta^\beta}\Big(\mathbf{T}^\beta + n^\beta \lagrange^{\beta} \mathbf{I}\Big)\frobenius\mathbf{L}_\beta \label{equation:entropy-fluid} \\
& + (n^F)^\prime_S \Big(s^M \lagrange^{M} \!+ s^C\lagrange^{C} \!- \lagrange^{S}\Big) \vphantom{\sum_\beta \frac{\theta^S}{\theta^\beta}} \label{equation:entropy-dalton} \\
& + (s^M)^\prime_S \Big(n^F \lagrange^{M} \!\!- n^F \lagrange^{C}  - \rho^M \frac{\partial \psi^M}{\partial s^M}\Big) \vphantom{\sum_\beta \frac{\theta^S}{\theta^\beta}} \label{equation:entropy-capillary} \\
& - \frac{\theta^S}{\theta^M}\Big(\hat{\mathbf{p}}^M\!\!-\lagrange^{M}\! \mathrm{grad}\,n^M \nonumber \\
& + \rho^M \frac{\partial \psi^M}{\partial s^M} \mathrm{grad}\,s^M\Big)\cdot\mathbf{w}_M \vphantom{\sum_\beta \frac{\theta^S}{\theta^\beta}} \label{equation:entropy-mfilter} \\
& - \frac{\theta^S}{\theta^C}\Big(\hat{\mathbf{p}}^C-\lagrange^{C} \mathrm{grad}\,n^C\Big)\cdot\mathbf{w}_C \vphantom{\sum_\beta \frac{\theta^S}{\theta^\beta}} \label{equation:entropy-cfilter} \\
& - \sum_{\alpha} \frac{\theta^S}{\theta^{\alpha}}\frac{1}{\theta^\alpha}\mathbf{q}^\alpha\cdot\mathrm{grad}\,\theta^\alpha \vphantom{\sum_\beta \frac{\theta^S}{\theta^\beta}} \label{equation:entropy-heatconduction} \\
& - \sum_{\alpha} \rho^\alpha\frac{\theta^S}{\theta^\alpha}\Big(\eta^\alpha+\frac{\partial\psi^\alpha}{\partial\theta^\alpha}\Big)(\theta^\alpha)^\prime_\alpha \vphantom{\sum_\beta \frac{\theta^S}{\theta^\beta}} \label{equation:entropy-ftr} \\
& - \sum_\alpha \rho^\alpha\frac{\theta^S}{\theta^\alpha}\Big( \lagrange^\alpha + \frac{\partial \psi^\alpha}{\partial v^{\alpha\mathrm{R}}} \Big) (v^{\alpha\mathrm{R}})^\prime_\alpha \label{equation:entropy-pressure} \\
& + \sum_\beta \Big( \hat{\varepsilon}^\beta\! +  \hat{\mathbf{p}}^\beta\!\!\cdot\! \mathbf{w}_\beta + \lagrange^\beta (n^\beta)^\prime_S \Big)\left(\frac{\theta^S}{\theta^\beta}-1\right) \nonumber \\
& -\rho^M \frac{\partial \psi^M}{\partial s^M} (s^M)^\prime_S \left(\frac{\theta^S}{\theta^M}-1\right)  \geq 0. \label{equation:entropy-heatexchange}
\end{align}
\label{equation:entropy-full}
\end{subequations}
The inequality~\eqref{equation:entropy-full} has to be fulfilled at all times in order for our model to be thermodynamically consistent, provided none of the preceding assumptions are violated or contradictory.\\
Our entropy inequality differs from typical formulations since we do not explicitly introduce the saturation constraint using the method of Lagrange multipliers~\cite{ehlers2009,Graf08}. However, it is necessary to employ the saturation constraint to arrive at our formulation, such that it is implicitly accounted for. In addition to that, the part~\eqref{equation:entropy-heatexchange} differs from some formulations under LTNE conditions (compare e.g.~\cite{Graf08}). To arrive at our formulation, we gathered the terms 
\begin{equation}
\Big(\hat{\mathbf{p}}^\beta\cdot \mathbf{w}_\beta + \lagrange^\beta (n^\beta)^\prime_S\Big) \left(\frac{\theta^S}{\theta^\beta} -1\right)
\end{equation}
and
\begin{equation}
-\rho^M \frac{\partial \psi^M}{\partial s^M} (s^M)^\prime_S \left(\frac{\theta^S}{\theta^M}-1\right)
\end{equation}
from the parts~\eqref{equation:entropy-dalton},~\eqref{equation:entropy-capillary},~\eqref{equation:entropy-mfilter} and~\eqref{equation:entropy-cfilter} and transferred them to the part~\eqref{equation:entropy-heatexchange}. Arising consequences are pointed out in the following.

\subsection{Coleman and Noll Procedure}\label{sec:coleman}
Following the Coleman and Noll procedure (cf.~\cite{Ehlers02}), each individual part of the entropy inequality is required to be identical to or greater than or equal to zero in the following.\\
For ease of notation, inspecting the parts~(\ref{equation:entropy-fluid}) and~(\ref{equation:entropy-dalton}), we identify the Lagrange multipliers $\lagrange^\alpha$ with the partial fluid pressures and the overall fluid pressure
\begin{equation}
\lagrange^\beta = p^{\beta\mathrm{R}}\quad\text{and}\quad \lagrange^S = p^{F\mathrm{R}}.
\end{equation}
Wherever it is convenient, these pressures are substituted already in the following.
\subsubsection{Solid Stress}\label{subsec:solidstress}
For the first part of the entropy inequality~(\ref{equation:entropy-solid}), we require
\begin{equation}\label{equation:solid-equilibrium}
\Big(\mathbf{T}^S_\mathrm{E}-\rho^S\frac{\partial\psi^S}{\partial \mathbf{F}_S}\mathbf{F}_S^\transpose\Big)\frobenius\mathbf{L}_{S} = 0
\end{equation}
with
\begin{equation}
\mathbf{T}^S_\mathrm{E} := \mathbf{T}^S+n^S p^{F\mathrm{R}}\,\mathbf{I}.
\end{equation}
Therein, we identify the solid extra stress $\mathbf{T}^S_\mathrm{E}$. Since the velocity gradient may assume arbitrary values, the terms in parentheses have to vanish. This can be ensured by modelling the solid constituent as hyperelastic; i.e., we assume
\begin{equation}\label{eq:hyperelastic}
\mathbf{T}^S_{\mathrm{E}} = J_S^{-1}\,\frac{\partial W^S}{\partial \mathbf{F}_S}\,\mathbf{F}_S^\transpose,\quad\mathrm{with}\quad J_S = \frac{\rho_{0S}^S}{\rho^S}.
\end{equation}
Therein, $W^S$ denotes a solid strain energy density function. An appropriate choice for $W^S$ is given in section~\ref{subsec:strainenergy}. Thereafter, the solid Cauchy stress tensor is recovered as
\begin{equation}\label{eq:solidstress}
\mathbf{T}^S = \mathbf{T}^S_\mathrm{E} - n^S p^{F\mathrm{R}}\,\mathbf{I}.
\end{equation}

\subsubsection{Fluid Stress}
For the second part of the entropy inequality~(\ref{equation:entropy-fluid}), we require
\begin{equation}\label{equation:fluid-equilibrium}
\mathbf{T}^\beta_\mathrm{E}\frobenius\mathbf{L}_\beta = 0,\quad\mathrm{with}\quad \mathbf{T}^\beta_\mathrm{E}:= \mathbf{T}^\beta + n^\beta p^{\beta\mathrm{R}}\,\mathbf{I}.
\end{equation}
Therein, we identify the fluid extra stresses $\mathbf{T}^\beta_\mathrm{E}$. Since the fluid velocity gradients may assume arbitrary values, the fluid extra stresses have to vanish. Therefore, we recover the fluid Cauchy stress tensors as
\begin{equation}\label{eq:fluidstress}
\mathbf{T}^\beta = -n^\beta p^{\beta\mathrm{R}}\,\mathbf{I}.
\end{equation}
The above requirement implies the neglect of viscous dissipation. In agreement with fundamental derivations~\cite{Ehlers02}, the friction force ($\mathrm{div}\,\mathbf{T}^F_\mathrm{E}$) is assumed to be negligible compared to the contribution arising from the extra momentum production ($\hat{\mathbf{p}}^\beta_\mathrm{E}$). As such, fluid viscosities are incorporated later (see section~\ref{subsec:filterlaws}).

\subsubsection{Overall Fluid Pressure}
For the third part of the entropy inequality~(\ref{equation:entropy-dalton}), we require
\begin{equation}
(n^F)^\prime_S \Big(s^M p^{M\mathrm{R}} + s^C p^{C\mathrm{R}} - p^{F\mathrm{R}}\Big) = 0.
\end{equation}
Since the fluid volume fraction may vary arbitrarily, the terms in parentheses have to vanish. This yields
\begin{equation}\label{equation:ltnedalton}
p^{F\mathrm{R}} = \sum_\beta s^\beta\,p^{\beta\mathrm{R}}.
\end{equation}
We recognise this relation as Dalton's law of partial pressures and we identify $p^{F\mathrm{R}}$ as the overall fluid or pore pressure. This relation is known from typical models under LTE conditions~\cite{Ehlers02,ehlers2009}. Our derivation implies that the same relation must hold under LTNE conditions.

\subsubsection{Capillary Pressure}\label{subsec:capillarypressure}
For the fourth part of the entropy inequality~(\ref{equation:entropy-capillary}), we require
\begin{equation}\label{equation:pdif-condition}
(s^M)^\prime_S \Big(n^F p^{M\mathrm{R}} - n^F p^{C\mathrm{R}}
- \rho^M \frac{\partial \psi^M}{\partial s^M}\Big) = 0.
\end{equation}
Since the marrow saturation may vary arbitrarily, the terms in parentheses have to vanish. The terms in parentheses vanish if
\begin{equation}\label{equation:pdif-relation}
\frac{\partial \psi^M}{\partial s^M} = \frac{n^F}{\rho^M} \Big(p^{M\mathrm{R}} - p^{C\mathrm{R}}\Big).
\end{equation}
Therein, following examples of the TPM~\cite{ehlers2009,Graf08}, the difference of the fluid pressures is identified as the equilibrium capillary or differential pressure
\begin{equation}\label{equation:pdif-relation2}
p^\mathrm{cap} =: p^{C\mathrm{R}} - p^{M\mathrm{R}}.
\end{equation}
It is defined by an empirical model $p^\mathrm{cap}$, i.e. a pressure saturation relation. An appropriate choice is specified in section~\ref{subsec:constitutivemodels}.\\
We recognise the relations~\eqref{equation:pdif-relation} and~\eqref{equation:pdif-relation2} from typical models under LTE conditions~\cite{Ehlers02,ehlers2009}. Our derivation implies that the same relations apply under LTNE conditions.

\subsubsection{Filter Laws}\label{subsec:filterlaws}
For the fifth and sixth part of the entropy inequality~(\ref{equation:entropy-mfilter}) and~(\ref{equation:entropy-cfilter}), we require
\begin{equation}\label{equation:pce}
-\hat{\mathbf{p}}^\beta_\mathrm{E}\cdot \mathbf{w}_\beta \geq 0,
\end{equation}
with
\begin{equation}\label{equation:pce2}
\hat{\mathbf{p}}^M_\mathrm{E}\! := \hat{\mathbf{p}}^M\! -p^{M\mathrm{R}} \mathrm{grad}\,n^M\!
+ \rho^M \frac{\partial \psi^M}{\partial s^M} \mathrm{grad}\,s^M
\end{equation}
and
\begin{equation}\label{equation:pce3}
\hat{\mathbf{p}}^C_\mathrm{E}:= \hat{\mathbf{p}}^C -p^{C\mathrm{R}} \mathrm{grad}\,n^C.
\end{equation}
Therein, we identify the extra momentum production $\hat{\mathbf{p}}^\beta_\mathrm{E}$ of the fluids.
Since the seepage velocities may assume arbitrary values, we require
\begin{equation}
\hat{\mathbf{p}}^\beta_\mathrm{E} \propto -\mathbf{w}_\beta,
\end{equation}
admitting
\begin{equation}\label{equation:pbetae}
\hat{\mathbf{p}}^{\beta}_{\mathrm{E}} := -(n^{\beta})^2\,\mu^{\beta\mathrm{R}}\,(\kappa_r^{\beta}\,\mathbf{K}^S)^{-1}\,\mathbf{w}_{\beta}.
\end{equation}
Therein, $\mu^{\beta\mathrm{R}}$ denotes the effective dynamic fluid viscosity, $\kappa^\beta_r$ the relative permeability factors of the fluids, and $\mathbf{K}^S$ the intrinsic permeability tensor of the solid.\\
Combining the definitions~(\ref{equation:pce2}) and~(\ref{equation:pbetae}) with the momentum balance of the marrow phase, we derive
\begin{equation}\label{eq:mfilterlaw}
\begin{split}
  & n^M\mathbf{w}_M = -\frac{\kappa_r^M\,\mathbf{K}^S}{\mu^{M\mathrm{R}}}\Big[\mathrm{grad}\,p^{M\mathrm{R}}-\rho^{M\mathrm{R}}\,\mathbf{b}
  \\
  & - p^\mathrm{cap} \frac{1}{s^M}  \mathrm{grad}\,s^M \Big].
\end{split}
\end{equation}
Therein, the last term in brackets arises due to relation~(\ref{equation:pdif-relation}). Analogously, for the cement phase we derive
\begin{equation}\label{eq:cfilterlaw}
n^C\mathbf{w}_C = -\frac{\kappa_r^C\,\mathbf{K}^S}{\mu^{C\mathrm{R}}}\Big[\mathrm{grad}\,p^{C\mathrm{R}}-\rho^{C\mathrm{R}}\,\mathbf{b}\Big].
\end{equation}
The above relations are filter laws, where the left-hand side is recognised as the Darcy velocity of the respective fluid constituent. These filter laws are known from typical models under LTE conditions~\cite{Ehlers02,ehlers2009}. Our derivation implies that the same filter laws apply under LTNE conditions.\\
The filter laws as well as the extra momentum production of the wetting and non-wetting phase differ by a contribution of the capillary pressure since only $\psi^M$ is assumed to depend on the marrow saturation $s^M$.\\
In agreement with the capillary pressure model, empirical models need to be assigned to the relative permeability factors. An appropriate choice is specified in section~\ref{subsec:constitutivemodels}.\\
Biological materials show anisotropic permeability in many cases, for which a transverse-isotropic permeability formulation is needed, as developed in \cite{Ricken.2007g, Pierce.2013}. We neglect this for simplicity, assuming isotropic permeability such that $\mathbf{K}^S = k^S\,\mathbf{I}$.\\
Finally, the viscosity of bone cement as well as bone marrow is known to be non-Newtonian; shear-thinning, in particular. We have investigated the viscosity of bone cement in other works~\cite{Trivedi2022,trivedi2024rheological}. Nevertheless, we employ constant viscosities for simplicity.

\subsubsection{Heat Conduction}
For the seventh part of the entropy inequality~(\ref{equation:entropy-heatconduction}), we require
\begin{equation}
- \sum_{\alpha} \frac{\theta^S}{\theta^{\alpha}}\frac{1}{\theta^\alpha}\mathbf{q}^\alpha\cdot\mathrm{grad}\,\theta^\alpha \geq 0.
\end{equation}
Since the temperature gradients may assume arbitrary values, we individually require
\begin{equation}\label{eq:fourierreq}
\mathbf{q}^\alpha \propto -\mathrm{grad}\,\theta^\alpha,
\end{equation}
admitting
\begin{equation}\label{eq:fourierslaw}
\mathbf{q}^\alpha := -n^\alpha\,\kappa^\alpha\,\mathrm{grad}\,\theta^\alpha.
\end{equation}
Therein, $\kappa^\alpha$ denotes the thermal conductivity of constituent $\varphi^\alpha$; thus, we recognise the above relation as Fourier's law of heat conduction. We neglect thermal dispersion for simplicity. Further, we neglect thermal tortuosity since the thermal properties of the constituents we consider are all within the same order of magnitude.
\subsubsection{Entropy and Heat Capacity}
For the eighth part of the entropy inequality~(\ref{equation:entropy-ftr}), we require
\begin{equation}
- \sum_{\alpha} \rho^\alpha\frac{\theta^S}{\theta^\alpha}\Big(\eta^\alpha+\frac{\partial\psi^\alpha}{\partial\theta^\alpha}\Big)(\theta^\alpha)^\prime_\alpha = 0.
\end{equation}
Since the temperatures may vary arbitrarily, the terms in parentheses have to vanish. We require
\begin{equation}\label{equation:cond3}
\eta^{\alpha} = - \frac{\partial \psi^{\alpha}}{\partial \theta^\alpha}.
\end{equation}
We recognise this as the thermodynamic definition of entropy, corresponding to the fundamental thermodynamic relation.\\
The entropy and the Helmholtz potential can be related to the specific heat capacity under constant generalised thermodynamic strain ($\mathbf{Z}^\alpha$), as defined by
\begin{equation}\label{eq:entropy4}
c_\mathbf{Z}^{\alpha} := \frac{\partial \varepsilon^{\alpha}}{\partial \theta^\alpha} = \theta^\alpha\frac{\partial \eta^{\alpha}}{\partial \theta^\alpha} = -\theta^\alpha\frac{\partial^2 \psi^{\alpha}}{\partial \theta^\alpha\partial \theta^\alpha}.
\end{equation}
Therein, we exploited the Legendre transformation of the internal energy~(\ref{equation:legendre1}) and relation~(\ref{equation:cond3}).
The generalised thermodynamic strain is representative of the material densities, volume fractions and deformation gradients which $\psi^\alpha$ is assumed to depend upon. Among these quantities, assuming only specific volumes to be temperature-dependent, we identify $c^\alpha_\mathbf{Z}$ as the isochoric heat capacity $c_v^\alpha$.\\

\subsubsection{Thermodynamic Pressure}
For the ninth part of the entropy inequality~(\ref{equation:entropy-pressure}), we require
\begin{equation}\label{eq:thermopressure}
- \sum_\alpha \rho^\alpha\frac{\theta^S}{\theta^\alpha}\Big( \lagrange^\alpha + \frac{\partial \psi^\alpha}{\partial v^{\alpha\mathrm{R}}} \Big) (v^{\alpha\mathrm{R}})^\prime_\alpha = 0.
\end{equation}
If material compressibility were to be assumed, the specific volumes could vary arbitrarily. Then, the terms in parentheses would need to vanish yielding the requirement
\begin{equation}\label{eq:thermodynamicpressure}
\lagrange^\alpha = - \frac{\partial \psi^\alpha}{\partial v^{\alpha\mathrm{R}}}.
\end{equation}
We recognise this as the thermodynamic definition of pressure, analogous to relation~(\ref{eq:thermodynamicpressure0}) and similar to relation~(\ref{equation:cond3}).

\subsubsection{Heat Transfer}
\label{subsec:heattransfer}
For the tenth part of the entropy inequality~(\ref{equation:entropy-heatexchange}), we require
\begin{equation}\label{equation:heatexchange-requirement}
\sum_\beta \frac{\hat{\varepsilon}_\mathrm{E}^\beta}{\theta^\beta}(\theta^S-\theta^\beta) \geq 0
\end{equation}
with
\begin{equation}\label{eq:extraenergyproduction}
\hat{\varepsilon}_\mathrm{E}^M\! := \hat{\varepsilon}^M\! +  \hat{\mathbf{p}}^M\!\!\cdot\! \mathbf{w}_M + \lagrange^M\! (n^M)^\prime_S + n^F\! p^\mathrm{cap} (s^M)^\prime_S
\end{equation}
and
\begin{equation}\label{eq:extraenergyproduction2}
\hat{\varepsilon}_\mathrm{E}^C := \hat{\varepsilon}^C +  \hat{\mathbf{p}}^C\!\!\cdot\! \mathbf{w}_C + \lagrange^C (n^C)^\prime_S.
\end{equation}
Therein, we define the extra energy production $\hat{\varepsilon}_\mathrm{E}^\beta$ of the fluids, where we employed relation~\eqref{equation:pdif-relation}. Since the constituent temperatures may vary arbitrarily, we may require
\begin{equation}
\hat{\varepsilon}_\mathrm{E}^\beta \propto (\theta^S - \theta^\beta),
\end{equation}
admitting
\begin{equation}
\hat{\varepsilon}_\mathrm{E}^\beta := \kappa^{S\beta}a^{S\beta}\,(\theta^S - \theta^\beta).
\end{equation}
Therein, $\kappa^{S\beta}$ denotes an interface-specific heat exchange coefficient, specific to the interface of the solid constituent and constituent $\varphi^\beta$. The coefficient $a^{S\beta}$ denotes the interface-area per unit volume for the same interface. Given these coefficients it is apparent that direct heat exchange between the fluid constituents is not considered using this approach. Instead, a more general approach is given by
\begin{equation}\label{eq:energyproduction}
\begin{split}
& \hat{\varepsilon}_\mathrm{E}^\beta := \sum_{\alpha \neq \beta} \kappa^{\alpha\beta}a^{\alpha\beta}(\theta^\alpha - \theta^\beta),
\\
& \mathrm{with}\quad \kappa^{\alpha\beta}a^{\alpha\beta} = \kappa^{\beta\alpha}a^{\beta\alpha} \geq 0.
\end{split}
\end{equation}
One may prove that with definition~(\ref{eq:energyproduction}), the requirement~(\ref{equation:heatexchange-requirement}) is fulfilled at all times~\cite{Graf08}. Appropriate constitutive relations for the above coefficients are given in section~\ref{subsec:poregeometry}.\\
By combining the momentum and energy production constraints~(\ref{eq:momenumconstr}), the direct energy production of the solid is recovered as
\begin{equation}\label{eq:solidenergyproduction}
\hat{\varepsilon}^S = -\sum_\beta \left(\hat{\varepsilon}^\beta +\hat{\mathbf{p}}^\beta\cdot\mathbf{w}_\beta\right).
\end{equation}
For ease of notation, we define the extra energy production of the solid as
\begin{equation}\label{eq:extrasolidenergyproduction}
\hat{\varepsilon}^S_\mathrm{E} := -\sum_\beta \hat{\varepsilon}^\beta_\mathrm{E}.
\end{equation}

\subsection{Constitutive Models}
In the following, the mentioned but yet to be specified constitutive models are given. Parameter values are given in Table~\ref{tab:modelparam}.

\begin{table}[htbp]
\begin{center}
\begin{minipage}{205pt}
\caption{Model parameters and references their values are taken from (see) or derived from (cf.)}\label{tab:modelparam}%
\begin{tabular}{@{}llll@{}}
\toprule
Symbol & Value & Unit & Reference\\
\midrule
$\rho^{S\mathrm{R}}$ & $1850.0$ & [kg/m$^3$] & see \cite{Keaveny04} \\
$\rho^{M\mathrm{R}}$ & $1060.0$ & [kg/m$^3$] & see \cite{Gurkan08} \\
$\rho^{C\mathrm{R}}$ & $1500.0$ & [kg/m$^3$] & cf. \cite{Kolmeder16} \\
$\mu^S$ & $3.85\times10^{9}$ & [Pa] & see~\cite{Trivedi2022} \\
$\lambda^S$ & $5.77\times10^{9}$& [Pa] & see~\cite{Trivedi2022} \\
$\mathbf{b}$ & $\mathbf{0.0}$ & [m/s$^2$] & arbitrary \\
$k^S$ & $5.0\times10^{\text{-}8}$ & [m$^2$] & cf. \cite{Kolmeder16} \\
$p_\mathrm{bc}$ & $1.0\times10^{\text{-}10}$ & [Pa] & arbitrary \\
$\lambda_\mathrm{bc}$ & $3.0$ & [ - ] & arbitrary \\
$s^M_\mathrm{res}$ & $0.00025$ & [ - ] & arbitrary \\
$s^C_\mathrm{res}$ & $0.00025$ & [ - ] & arbitrary \\
$\mu_{M\mathrm{R}}$ & $50.0$ & [Pa$\,$s] & arbitrary \\
$\mu_{C\mathrm{R}}$ & $50.0$ & [Pa$\,$s] & arbitrary \\
$c_p^S$ & $2274.0$ & [J/(kg$\,$K)] & see \cite{McIntosh10} \\
$c_p^M$ & $2666.0$ & [J/(kg$\,$K)] & see \cite{McIntosh10} \\
$c_p^C$ & $1470.0$ & [J/(kg$\,$K)] & cf. \cite{Kolmeder16} \\
$\kappa^S$ & $0.42$ & [W/(m$\,$K)] & see \cite{Feldmann18} \\
$\kappa^M$ & $0.42$ & [W/(m$\,$K)] & cf. \cite{Feldmann18} \\
$\kappa^C$ & $0.25$ & [W/(m$\,$K)] & cf. \cite{Kolmeder16} \\
$r^\alpha$ & $0.0$ & [W/kg] & arbitrary \\
$L^F$ & $1.0\times10^{\text{-}3}$ & [m] & cf. \cite{Keaveny01,Keaveny04} \\
$q^F$ & $0.6$ & [ - ] & cf. \cite{coelho2009numerical} \\
\botrule
\end{tabular}
\end{minipage}
\end{center}
\end{table}

\subsubsection{Solid Strain Energy Density}\label{subsec:strainenergy}
To model the mechanical behaviour of the solid constituent, a strain energy density function $W^S$ has to be specified. Assuming isotropic material behaviour, a neo-Hookean model with a volumetric extension term~(e.g.~\cite{Bleiler15}) is given by
\begin{equation}
\begin{split}
& W^S := {\textstyle\frac{1}{2}}\, \mu^S (\mathrm{tr}(\mathbf{C}_S) - 3)- \mu^S \mathrm{ln}\,J_S
\\
& + \lambda^S\!\left(1-n_{0S}^S\right)^2\! \left(\frac{J_S-1}{1-n^S_{0S}}-\mathrm{ln}\frac{J_S-n^S_{0S}}{1-n^S_{0S}}\right).
\end{split}
\end{equation}
Therein, the first two terms correspond to the standard formulation of the neo-Hookean model for solid materials, whereas the remaining term accounts for the fact that, within the TPM, the compaction point is given by the initial volume fraction $n^S_{0S}$. This yields the solid extra stress
\begin{equation}\label{eq:neohookean}
\begin{split}
& \mathbf{T}^S_{\mathrm{E}} = \frac{\mu^S}{J_S}\,\left(\mathbf{F}_S\,\mathbf{F}_S^\transpose-\mathbf{I}\right)
\\
& +\lambda^S\,\left(1-n_{0S}^S\right)\,\left(\frac{J_S-1}{J_S-n_{0S}^S}\right)\,\mathbf{I}.
\end{split}
\end{equation}
Therein, $\mu^S$ and $\lambda^S$ are the first and second Lam\'{e} constants of the solid constituent.\\
For our application, solid deformations are typically negligible~\cite{Trivedi2022}. As such, we may linearise the above relations. Anticipating the numerical treatment, we omit this.

\subsubsection{Capillary Pressure and Relative Permeability}\label{subsec:constitutivemodels}
In our application, cement is the non-wetting and bone marrow the wetting fluid. As such, the cement injection we consider is a drainage process. An appropriate capillary pressure saturation relation for the description of drainage processes is the Brooks-Corey model~\cite{lenhard1989correspondence} given by
\begin{equation}\label{eq:capillarypressure}
p^\mathrm{cap}:= p_\mathrm{bc}\,(s^M_\mathrm{eff})^{-1/\lambda_\mathrm{bc}},
\end{equation}
with
\begin{equation}
s^M_\mathrm{eff}:= \frac{s^M - s^M_\mathrm{res}}{1 - s^M_\mathrm{res} - s^C_\mathrm{res}}.
\end{equation}
Therein, $p_\mathrm{bc}$ and $\lambda_\mathrm{bc}$ are the entry pressure and uniformity parameter of the Brooks-Corey model. Further, an effective bone marrow saturation is introduced, for which residual cement and marrow saturations are necessary parameters.\\
The models for the relative permeability factors are closely related to the model for the capillary pressure. The Brooks-Corey relative permeability factors~\cite{lenhard1989correspondence} are given as
\begin{equation}\label{eq:krelm}
\kappa_r^M := (s^M_\mathrm{eff})^{(2+3 \lambda_\mathrm{bc})/\lambda_\mathrm{bc}}
\end{equation}
and
\begin{equation}\label{eq:krelc}
\kappa_r^C := (1-s^M_\mathrm{eff})^2\left[1-(s^M_\mathrm{eff})^{(2+\lambda_\mathrm{bc})/\lambda_\mathrm{bc}}\right],
\end{equation}
employing the same uniformity parameter as the model for the capillary pressure. Other capillary pressure saturation formulations within the TPM have been formulated, e.g. in \cite{Ricken.2003c}.\\
For our application, capillary forces can be expected to have negligible influence~\cite{Trivedi2022}. In accordance, we employ a value of $p_\mathrm{bc}$ which renders $p^\mathrm{cap}$ negligible. Under these conditions, the relative permeability factors may be modelled as linearly saturation-dependent~\cite{avraam1995flow}, as given by
\begin{equation}
\kappa^M_r := s^M_\mathrm{eff} \quad\text{and}\quad \kappa^C_r := 1 - s^M_\mathrm{eff}.
\end{equation}

\subsubsection{Heat Exchange Coefficients and Interface-Area}\label{subsec:poregeometry}
\begin{figure}[htbp]
\includegraphics[scale=0.25]{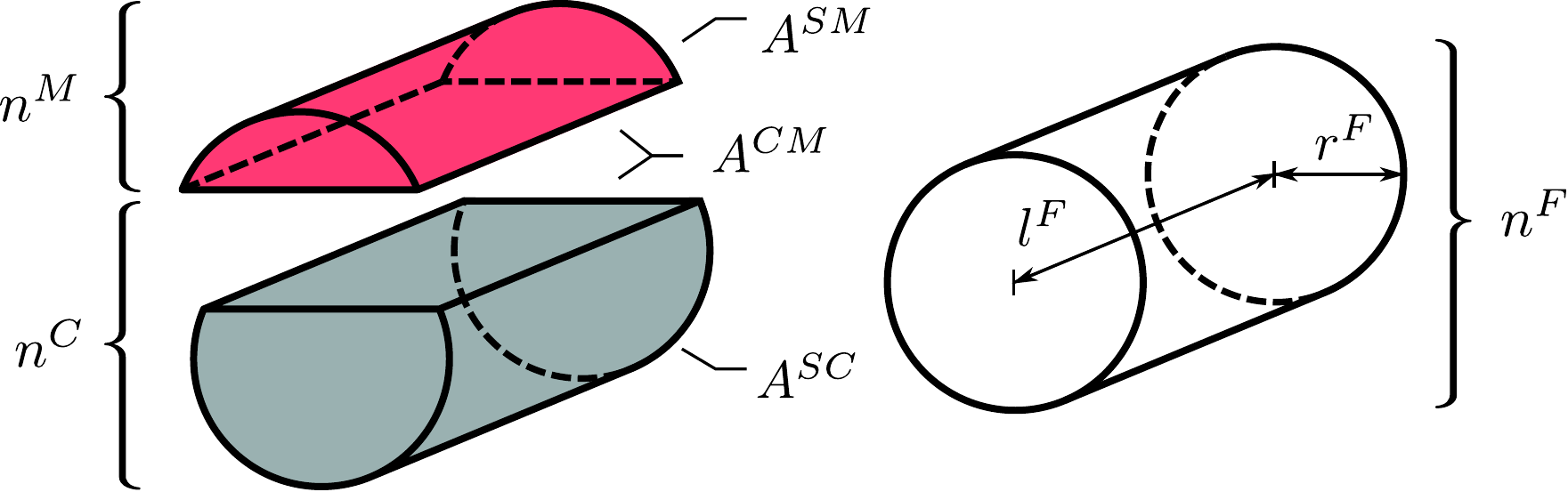}
\caption{Dimensionless pore geometry. See text for description}
\label{fig:poregeometry}
\end{figure}

\begin{figure}[htbp]
\includegraphics[scale=1.0]{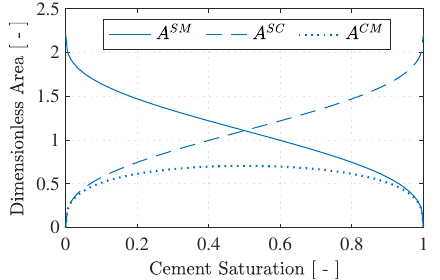}
\caption{Plots of dimensionless interface area against cement saturation for $q^F=0.6$ and $n^F=0.85$}
\label{fig:dimlessarea}
\end{figure}

Constitutive assumptions for the heat exchange coefficients $\kappa^{\alpha\beta}$ and the interface-areas per unit volume $a^{\alpha\beta}$ have to be made.\\
The transfer of heat shall be based on heat conductance only. Convective heat transfer in multiphase porous media flow is part of current research and neglected for simplicity (cf.~\cite{nuske2014,nuske2015}).\\
As such, the heat exchange coefficients are computed as
\begin{equation}
\kappa^{\alpha\beta} = \frac{1}{L^F}\,\Big(\frac{1}{\kappa^\alpha} + \frac{1}{\kappa^\beta}\Big)^{-1}.
\end{equation}
Therein, a reference length $L^F$ is introduced. We choose it based on the average intertrabecular spacing of trabecular bone.\\
To define the interface-area per unit volume, we propose a pore geometry from which interface-areas per unit volume can be determined for given fluid volume fractions.\\
Let a normalised pore be given by an open tube with volume $n^F$, as depicted in Figure~\ref{fig:poregeometry}. The radius and length of the tube are denoted $r^F$ and $l^F$, respectively. The ratio $q^F:=l^F/r^F$ shall be a given material parameter. Then, $r^F$ and $l^F$ are uniquely determined by the tube's volume. The surface of the tube represents the interface of the solid constituent and the overall fluid constituent. The area of the circular bases is not part of this interface.\\
As depicted, the tube is split longitudinally, into a tubular section with volume $n^C$ and a tubular section with volume $n^M$. The shared interface of these tubular sections is the interface of the fluid phases with area $A^{CM}$. The remaining surface of either tubular section is the interface of the respective fluid phase and the solid phase with area $A^{S\beta}$. The dependence on the cement saturation is visualised in Figure~\ref{fig:dimlessarea}.\\
By definition, the volume fractions are dimensionless, hence, the interface-areas determined from the above geometry are dimensionless as well. Division by the reference length $L^F$ yields the solid-fluid interface area per unit volume $a^{S\beta}:=A^{S\beta}/L^F$. The fluid-fluid interface area $a^{CM}:=A^{CM}$ is assumed to scale proportionally to the total fluid volume, omitting the division by~$L^F$. The model parameters are chosen such that the solid-fluid interface area ($a^{SM}+a^{SC}$) corresponds to the specific surface area of trabecular bone (see e.g.~\cite{coelho2009numerical} and Fig.~\ref{fig:dimlessarea}).

\subsection{Further Considerations}\label{subsec:inconsistent}
In the following, we reformulate the constituent energy balances as heat transport equations.

\subsubsection{Derivatives of the Entropies}
Anticipating the next subsection, we need to consider the material time derivatives of the entropies and, thus, the following second-order derivatives.\\
Exploiting the symmetry of second-order derivatives as well as the relations~\eqref{eq:hyperelastic} and \eqref{eq:neohookean}, results in
\begin{equation}
\frac{\partial \eta^S}{\partial \mathbf{F}_S} = -\frac{\partial}{\partial \theta^S}\left(\frac{\partial \psi^S}{\partial \mathbf{F}_S} \right) = \mathbf{0}.
\end{equation}
Exploiting the relations~\eqref{equation:pdif-relation} and \eqref{eq:capillarypressure}, provides
\begin{equation}\label{eq:etacapillary}
\frac{\partial \eta^M}{\partial s^M} = -\frac{\partial}{\partial \theta^M}\left(\frac{\partial \psi^M}{\partial s^M} \right) = 0.
\end{equation}
Exploiting the relation~\eqref{eq:thermodynamicpressure}, yields
\begin{equation}
\frac{\partial \eta^\alpha}{\partial v^{\alpha\mathrm{R}}} = -\frac{\partial}{\partial \theta^\alpha}\left(\frac{\partial \psi^\alpha}{\partial v^{\alpha\mathrm{R}}} \right) = \frac{\partial \lagrange^\alpha}{\partial \theta^\alpha}.
\end{equation}
Exploiting the relations~\eqref{equation:cond3} and \eqref{eq:entropy4}, gives
\begin{equation}
\frac{\partial \eta^\alpha}{\partial \theta^\alpha} = -\frac{\partial}{\partial \theta^\alpha}\left(\frac{\partial \psi^\alpha}{\partial \theta^\alpha} \right) = \frac{c_v^\alpha}{\theta^\alpha}.
\end{equation}

\subsubsection{Constituent Energy Balances}\label{subsec:heuristicenergy}
The constituent energy balances~(\ref{eq:energy}) can be rewritten by employing the Legendre transformation of the internal energy~(\ref{equation:legendre1}) such that
\begin{equation}\label{eq:energy1}
\begin{split}
&\rho^\alpha \left[(\psi^\alpha)^\prime_\alpha + \eta^\alpha (\theta^\alpha)^\prime_\alpha + \theta^\alpha (\eta^\alpha)^\prime_\alpha\right] \\
& = \mathbf{T}^\alpha \frobenius \mathbf{L}_\alpha - \mathrm{div}\,\mathbf{q}^\alpha + \rho^\alpha r^\alpha + \hat{\varepsilon}^\alpha.
\end{split}
\end{equation}
On the left-hand side, the material time derivatives expand according to the chain rule of differentiation as
\begin{equation}
(\psi^\alpha)^\prime_\alpha = \frac{\partial \psi^\alpha}{\partial \theta^\alpha} (\theta^\alpha)^\prime_\alpha + \frac{\partial \psi^\alpha}{\partial v^{\alpha\mathrm{R}}} (v^{\alpha\mathrm{R}})^\prime_\alpha + \frac{\partial \psi^\alpha}{\partial \xi^\alpha} (\xi^\alpha)^\prime_\alpha
\end{equation}
and
\begin{equation}
(\eta^\alpha)^\prime_\alpha = \frac{\partial \eta^\alpha}{\partial \theta^\alpha} (\theta^\alpha)^\prime_\alpha + \frac{\partial \eta^\alpha}{\partial v^{\alpha\mathrm{R}}} (v^{\alpha\mathrm{R}})^\prime_\alpha + \frac{\partial \eta^\alpha}{\partial \xi^\alpha} (\xi^\alpha)^\prime_\alpha,
\end{equation}
where $\xi^\alpha$ is representative for $\mathbf{F}_S$ and $s^M$, respectively. On the right-hand side of equation~\eqref{eq:energy1}, we express the stress-power as
\begin{equation}
\mathbf{T}^S\!\!\frobenius\! \mathbf{L}_S = \mathbf{T}^S_\mathrm{E} \!\frobenius\! \mathbf{L}_S + \lagrange^S (n^S)^\prime_S - \rho^S \lagrange^S (v^{S\mathrm{R}})^\prime_S
\end{equation}
or
\begin{equation}
\mathbf{T}^\beta\frobenius \mathbf{L}_\beta = \lagrange^\beta (n^\beta)^\prime_\beta - \rho^\beta \lagrange^\beta (v^{\beta\mathrm{R}})^\prime_\beta,
\end{equation}
respectively. Using the relations~\eqref{eq:hyperelastic} and  \eqref{eq:neohookean}, provides
\begin{equation}
\rho^S \frac{\partial \psi^S}{\partial \mathbf{F}_S} \frobenius (\mathbf{F}_S)^\prime_S = \mathbf{T}^S_\mathrm{E} \frobenius \mathbf{L}_S.
\end{equation}
Using the relations~\eqref{equation:pdif-relation} and \eqref{equation:pdif-relation2}, results in
\begin{equation}
\rho^M \frac{\partial \psi^M}{\partial s^M} (s^M)^\prime_M = -n^F p^\mathrm{cap} (s^M)^\prime_M.
\end{equation}
Using the relation~\eqref{eq:thermodynamicpressure}, provides
\begin{equation}
\rho^\alpha \frac{\partial \psi^\alpha}{\partial v^{\alpha\mathrm{R}}} (v^{\alpha\mathrm{R}})^\prime_\alpha = - \rho^\alpha \lagrange^\alpha (v^{\alpha\mathrm{R}})^\prime_\alpha.
\end{equation}
Using the relation~\eqref{equation:cond3}, yields
\begin{equation}
\rho^\alpha \frac{\partial \psi^\alpha}{\partial \theta^\alpha} (\theta^\alpha)^\prime_\alpha + \rho^\alpha \eta^\alpha (\theta^\alpha) = 0.
\end{equation}
We recognise the terms in the above identities as part of the constituent energy balances. Exploiting the identities, the energy balances reduce to
\begin{equation}\label{eq:energy2}
\begin{split}
& \rho^\alpha c_v^\alpha (\theta^\alpha)^\prime_\alpha + \rho^\alpha \theta^\alpha \frac{\partial \lagrange^\alpha}{\partial \theta^\alpha} (v^{\alpha\mathrm{R}})^\prime_\alpha \\
& = - \mathrm{div}\,\mathbf{q}^\alpha + \rho^\alpha r^\alpha  + \omega^\alpha + \hat{\varepsilon}^\alpha_\mathrm{E}.
\end{split}
\end{equation}
On the right-hand side, we introduced dissipation terms $\omega^\alpha$ as
\begin{equation}
\omega^M \! := \lagrange^M (n^M)^\prime_M + n^F p^\mathrm{cap} (s^M)^\prime_M + \hat{\varepsilon}^M\! - \hat{\varepsilon}^M_\mathrm{E}
\end{equation}
and else
\begin{equation}
\omega^\alpha := \lagrange^\alpha (n^\alpha)^\prime_\alpha + \hat{\varepsilon}^\alpha - \hat{\varepsilon}^\alpha_\mathrm{E}.
\end{equation}
Using the relations~\eqref{eq:extraenergyproduction},~\eqref{eq:solidenergyproduction} and~\eqref{eq:extrasolidenergyproduction}, this is equivalent to
\begin{equation}
\omega^S\! = 0\!\quad\text{and}\quad\! \omega^\beta\! = \mu^{\beta\mathrm{R}}  (\kappa_r^{\beta}\,\mathbf{K}^S)^{-1}(n^\beta \mathbf{w}_\beta)^2\!.
\end{equation}
On the left-hand side of equation~\eqref{eq:energy2}, we may formally assume that the specific volumes depend on temperature. Then, the left hand-side becomes
\begin{equation}
\rho^\alpha\! \left[ c_v^\alpha + \theta^\alpha \frac{\partial \lagrange^\alpha}{\partial \theta^\alpha} \frac{\partial v^{\alpha\mathrm{R}}}{\partial \theta^\alpha} \right]\! (\theta^\alpha)^\prime_\alpha = \rho^\alpha c_p^\alpha (\theta^\alpha)^\prime_\alpha,
\end{equation}
where we recognise the formal definition of the isobaric specific heat capacities. Finally, we recover the constituent energy balances as
\begin{equation}
\rho^\alpha c_p^\alpha (\theta^\alpha)^\prime_\alpha = - \mathrm{div}\,\mathbf{q}^\alpha + \rho^\alpha r^\alpha + \omega^\alpha + \hat{\varepsilon}^\alpha_\mathrm{E}.
\end{equation}

\subsection{Numerical Treatment}
In this section, our model is prepared for numerical discretisation. The spatial discretisation is based on the Petrov-Galerkin finite element method. In particular, the Box discretisation is employed, a full upwinding approach (e.g.~\cite{hackbusch1989first,huber2000node}). The temporal discretisation is done employing a Crank-Nicholson scheme. The system of governing equations is solved monolithically using the coupled finite element solver PANDAS\footnote{\textbf{P}orous media \textbf{A}daptive \textbf{N}onlinear finite-element solver based on \textbf{D}ifferential \textbf{A}lgebraic \textbf{S}ystems (http://www.get-pandas.com)}.
\subsubsection{Governing Differential Equations}\label{subsec:governing}
The primary variables are chosen as $\mathbf{u}_S$, $s^{M}$, $p^{C\mathrm{R}}$, $\theta^S$, $\theta^M$ and $\theta^C$; where $\mathbf{u}_S$ has three dimensions.\\
Analogous to the isothermal case (cf.~\cite{Bleiler15}), three governing equations are given by the aggregate momentum balance
\begin{equation}
\mathbf{0} = \underbrace{\mathrm{div}\Big(\sum_\alpha \mathbf{T}^\alpha\Big)}_{\displaystyle =\mathrm{div}\,\mathbf{T}} + \underbrace{\Big(\sum_\alpha \rho^\alpha\Big)}_{\displaystyle =:\rho}\mathbf{b}
\end{equation}
and two by the fluid constituent volume balances
\begin{equation}\label{equation:strongvolume1}
0 = (n^{\beta})^{\prime}_S + \mathrm{div}(n^{\beta}\,\mathbf{w}_{\beta}) + n^{\beta}\,\mathrm{div}\,(\mathbf{u}_S)^{\prime}_S.
\end{equation}
Under LTNE conditions, this equation system is expanded by the solid energy balance
\begin{equation}
0 = \rho^S c_p^S\, (\theta^S)^\prime_S + \mathrm{div}\,\mathbf{q}^S - \rho^S r^S - \hat{\varepsilon}^S_\mathrm{E}
\end{equation}
and the fluid energy balances
\begin{equation}
\begin{split}
& 0 = \rho^\beta c_p^\beta\, (\theta^\beta)^\prime_S + \rho^{\beta\mathrm{R}} c_p^\beta\mathrm{div}(\theta^\beta n^\beta\mathbf{w}_\beta)
\\
& + \rho^{\beta\mathrm{R}} c_p^\beta\theta^\beta \left[ (n^\beta)^\prime_S + n^\beta\,\mathrm{div}\,(\mathbf{u}_S)^\prime_S\right]
\\
& + \mathrm{div}\,\mathbf{q}^\beta - \rho^\beta r^\beta - \omega^\beta -\hat{\varepsilon}^\beta_\mathrm{E}.
\end{split}
\end{equation}

\subsubsection{Weak Formulations}
Denoting the not yet specified simulation domain as $\Omega$ and the test functions as $\delta\boldsymbol{\varphi}$ and $\delta\varphi$, the weak formulation of the momentum balance is derived as
\begin{equation}
\begin{split}
 & 0 = \int_{\Omega} \left( -\mathbf{T}\frobenius\mathrm{grad}\,\delta\boldsymbol{\varphi} + \rho\,\mathbf{b}\cdot\delta\boldsymbol{\varphi}\right)\,\mathrm{d}v
 \\
 & + \int_{\Gamma_{\mathbf{t}}} \delta\boldsymbol{\varphi}\cdot\mathbf{T}\mathbf{n}\,\mathrm{d}a,
\end{split}
\end{equation}
the weak formulation of the fluid volume balances is derived as
\begin{equation}
\begin{split}
  & 0 = \int_{\Omega} \Big( (n^{\beta})^{\prime}_S + n^{\beta}\,\mathrm{div}\,(\mathbf{u}_S)^{\prime}_S\Big)\,\delta \varphi\,\mathrm{d}v
  \\
  & - \int_{\Omega} n^{\beta}\,\mathbf{w}_{\beta}\cdot\mathrm{grad}\,\delta \varphi\,\mathrm{d}v
  \\
  & + \int_{\Gamma_{v^{\beta}}} \delta \varphi\,\underbrace{n^{\beta}\,\mathbf{w}_{\beta}\cdot\mathbf{n}}_{\displaystyle =:v^\beta}\,\mathrm{d}a,
\end{split}
\end{equation}
the weak formulation of the solid energy balance is derived as
\begin{equation}
\begin{split}
  & 0 =\int_{\Omega} \Big(\rho^S c_p^S (\theta^S)^\prime_S - \rho^S r^S\!\! - \hat{\varepsilon}^S_\mathrm{E} \Big)\delta\varphi\,\mathrm{d}v 
  \\
  & + \int_{\Omega} n^S\kappa^S\,\mathrm{grad}\,\theta^S\cdot\mathrm{grad}\,\delta\varphi\,\mathrm{d}v
  \\
  & - \int_{\Gamma_{\mathrm{q}^S}} \delta\varphi\,\underbrace{n^S \kappa^S\,\mathrm{grad}\,\theta^S\cdot\mathbf{n}}_{\displaystyle =:\mathrm{q}^S}\,\mathrm{d}a
\end{split}
\end{equation}
and the weak formulation of the fluid energy balances is derived as
\begin{equation}
\begin{split}
  & 0 = \int_{\Omega} \Big(\rho^\beta c_p^\beta\, (\theta^\beta)^\prime_S - \rho^\beta r^\beta - \omega^\beta -\hat{\varepsilon}^\beta_\mathrm{E}
  \\
  & + \rho^{\beta\mathrm{R}} c_p^\beta\theta^\beta \left[ (n^\beta)^\prime_S + n^\beta\,\mathrm{div}\,(\mathbf{u}_S)^\prime_S\right]\Big)\,\delta\varphi\,\mathrm{d}v 
  \\
  & - \int_{\Omega} \rho^{\beta\mathrm{R}} c_p^\beta \theta^\beta n^\beta \mathbf{w}_\beta \cdot \mathrm{grad}\,\delta\varphi\,\mathrm{d}v
  \\
  & + \int_{\Gamma_{v^\beta_\theta}} \delta\varphi\, \underbrace{\rho^{\beta\mathrm{R}} c_p^\beta \theta^\beta n^\beta \mathbf{w}_\beta \cdot \mathbf{n}}_{\displaystyle =:v^\beta_\theta}\,\mathrm{d}a
  \\
  & + \int_{\Omega} n^\beta\kappa^\beta\,\mathrm{grad}\,\theta^\beta\cdot\mathrm{grad}\,\delta\varphi\,\mathrm{d}v
  \\
  &- \int_{\Gamma_{\mathrm{q}^\beta}} \delta\varphi\,\underbrace{n^\beta \kappa^\beta\,\mathrm{grad}\,\theta^\beta\cdot\mathbf{n}}_{\displaystyle =:\mathrm{q}^\beta}\,\mathrm{d}a.
\end{split}
\end{equation}
Therein, as Neumann boundaries we identify the traction boundary $\Gamma_\mathbf{t}$, the volume flux boundaries $\Gamma_{v^\beta}$ as well as the heat flux boundaries $\Gamma_{\mathrm{q}^\alpha}$ and $\Gamma_{v^\beta_\theta}$ for conductive and advective heat flux, respectively.
\subsubsection{Geometry and Discretisation}
\begin{figure}[htbp]
\includegraphics[scale=0.22]{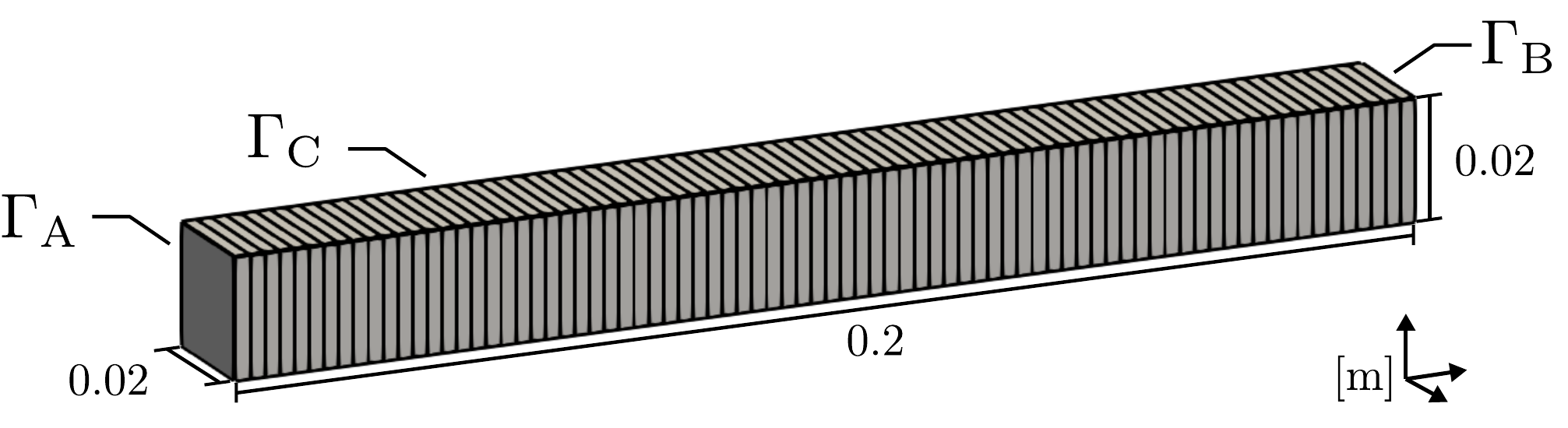}
\caption{Discretised simulation domain $\Omega$ with indicated boundaries $\Gamma_\mathrm{A}$, $\Gamma_\mathrm{B}$ and $\Gamma_\mathrm{C}$}
\label{fig:geometry}
\end{figure}

A simple tubular geometry with quadratic cross-section is considered, as depicted in Figure~\ref{fig:geometry}. Three boundaries are distinguished. Boundary $\Gamma_\mathrm{A}$ at one end of the tube, boundary $\Gamma_\mathrm{B}$ at the opposite end as well as $\Gamma_\mathrm{C}$, the mantle of the tube.\\
The finite element mesh consists of 80 hexahedral elements made up of 324 nodes. Linear shape functions are considered for all primary variables since this simplifies the numerical implementation considerably. For the employed Box approach the test functions are element-wise constant. Our simulations converge without quadratic discretisation of displacement since solid deformations are suppressed by construction. For the following scenarios all simulations are stable for at least 200 seconds with a fixed time step-size of $1.0$ seconds.  For a convergence study see appendix~\ref{subsec:convergence}.

\subsection{Simulation Scenarios}\label{sec:scenarios}
Two scenarios are simulated with our model in order to investigate and demonstrate its behaviour. All parameter values are given in Table~\ref{tab:modelparam}, unless explicitly stated otherwise.

\subsubsection{Scenario 1}
In scenario 1, cement is injected into the tube at boundary $\Gamma_\mathrm{A}$. The boundary $\Gamma_\mathrm{B}$ represents the outflow boundary. The inflowing cement has a temperature of $308.15\,\mathrm{K}$ which is identical to the initial temperature. The tube is insulated regarding heat conduction. All initial and boundary values are given in Table~\ref{tab:scenario1}.\\
Scenario 1 is simulated two times, corresponding to the following cases:
\begin{itemize}
\setlength{\itemindent}{5mm}
\item[{[1.a]}] Simulation with model employing Brooks-Corey relative permeability factors.
\item[{[1.b]}] Simulation with model employing linear relative permeability factors.
\end{itemize}

\subsubsection{Scenario 2}
In scenario 2, the inflowing cement has a temperature of $298.15\,\mathrm{K}$, which is below the initial temperature. Scenario 2 deviates from scenario 1 by the boundary condition given in Table~\ref{tab:scenario3} and is identical to scenario 1 otherwise.\\
Scenario 2 is simulated two times, corresponding to the following cases:
\begin{itemize}
\setlength{\itemindent}{5mm}
\item[{[2.a]}] Simulation with model employing Brooks-Corey relative permeability factors.
\item[{[2.b]}] Simulation with model employing linear relative permeability factors.
\end{itemize}

\begin{table}[htbp]
\begin{center}
\begin{minipage}{205pt}
\renewcommand*{\thefootnote}{\emph{\alph{footnote}}}
\caption{Initial and boundary conditions for scenario 1}\label{tab:scenario1}%
\begin{tabular}{@{}llll@{}}
\toprule
Symbol & Value & Unit & Domain/Boundary\\
\midrule
$\mathbf{u}_{0S}$ & $\mathbf{0.0}$ & [m] & $\Omega$ \\
$s^M_{0S}$ & $0.99975$ & [ - ] & $\Omega$ \\[1pt]
$p^{C\mathrm{R}}_{0S}$ & $0.0$ & [Pa] & $\Omega$ \\[1pt]
$\theta^\alpha_{0S}$ & $308.15$ & [K] & $\Omega$ \\[1pt]
$n^S_{0S}$ & $0.15$ & [ - ] & $\Omega$ \\
$\mathbf{u}_S$ & $\mathbf{0.0}$ & [m] & $\Gamma_\mathrm{A},\Gamma_\mathrm{B},\Gamma_\mathrm{C}$ \\[1pt]
$s^M$ & $0.95$ & [ - ] & $\Gamma_\mathrm{B}$ \\
$p^{C\mathrm{R}}$ & $0.0$ & [Pa] & $\Gamma_\mathrm{B}$ \\
$\theta^\alpha$ & $308.15$ & [K] & $\Gamma_\mathrm{B}$ \\
$v^M$ & $0.0$ & [m/s] & $\Gamma_\mathrm{A},\Gamma_\mathrm{C}$ \\
$v^C$ & $5.0\times10^{\text{-}4}$ & [m/s] & $\Gamma_\mathrm{A}$ \\
$v^C$ & $0.0$ & [m/s] & $\Gamma_\mathrm{C}$ \\
$\mathrm{q}^\alpha$ & $0.0$ & [W/m$^2$] & $\Gamma_\mathrm{A},\Gamma_\mathrm{B},\Gamma_\mathrm{C}$ \\[1pt]
$v^M_\theta$ & $0.0$ & [W/m$^2$] & $\Gamma_\mathrm{A},\Gamma_\mathrm{C}$ \\[1pt]
$v^C_\theta$ & $3.40\times10^5$ \footnotemark[1] & [W/m$^2$] & $\Gamma_\mathrm{A}$ \\[1pt]
$v^C_\theta$ & $0.0$ & [W/m$^2$] & $\Gamma_\mathrm{C}$ \\
\botrule
\end{tabular}
\footnotetext[]{\textsuperscript{\emph{a}}From $v^C_\theta = \rho^{C\mathrm{R}} c_p^C \theta^C_\mathrm{b.c.} v^C$ with $\theta^C_\mathrm{b.c.}=308.15\,\mathrm{K}$}
\end{minipage}
\end{center}
\end{table}

\begin{table}[htbp]
\begin{center}
\begin{minipage}{205pt}
\renewcommand*{\thefootnote}{\emph{\alph{footnote}}}
\caption{Boundary condition for scenario 2}\label{tab:scenario3}%
\begin{tabular}{@{}llll@{}}
\toprule
Symbol & Value & Unit & Domain/Boundary\\
\midrule
$v^C_\theta$ & $3.29\times10^5$ \footnotemark[1] & [W/m$^2$] & $\Gamma_\mathrm{A}$ \\
\botrule
\end{tabular}
\footnotetext[]{\textsuperscript{\emph{a}}From $v^C_\theta = \rho^{C\mathrm{R}} c_p^C \theta^C_\mathrm{b.c.} v^C$ with $\theta^C_\mathrm{b.c.}=298.15\,\mathrm{K}$}
\end{minipage}
\end{center}
\end{table}

\section{Results}
In the following, the results of the simulations of scenarios 1 and 2 are shown and discussed.\\
In all following figures, the values of selected variables are plotted along the main axis of the considered tubular geometry; i.e., we consider profiles along flow direction. They are shown at 30, 60, 90 and 120 seconds of simulated time, unless explicitly stated otherwise.\\
For all simulations, we prescribe zero displacement at the boundary of our simulation domain. Further, all nodes of the considered finite element mesh are boundary nodes. As such, by construction, solid deformations are zero at all times and we omit their visualisation in the following.
\subsection{Common Results}
\begin{figure}[htbp]
\includegraphics[scale=1.0]{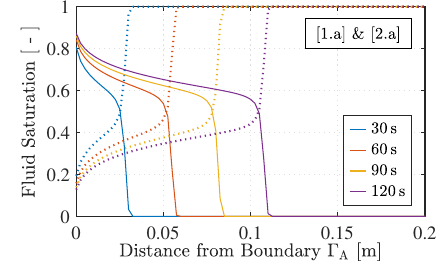}
\includegraphics[scale=1.0]{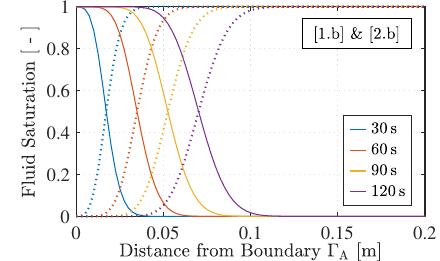}
\caption{Profiles of the cement (---) and marrow ($\cdots\:\!$) fluid saturation. Common results with Brooks-Corey (top) and linear relative permeability factors (bottom)}
\label{fig:results-saturation}
\end{figure}

\begin{figure}[htbp]
\includegraphics[scale=1.0]{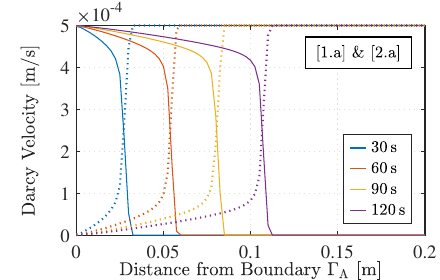}
\includegraphics[scale=1.0]{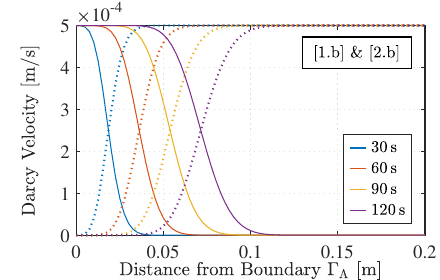}
\caption{Profiles of the cement (---) and marrow ($\cdots\:\!$) Darcy velocity. Common results with Brooks-Corey (top) and linear relative permeability factors (bottom)}
\label{fig:results-darcy}
\end{figure}

\begin{figure}[htbp]
\includegraphics[scale=1.0]{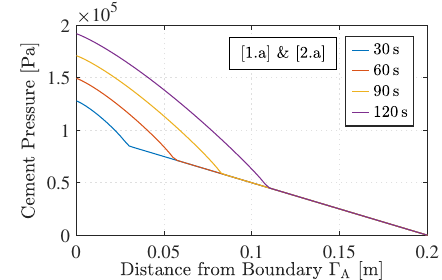}
\includegraphics[scale=1.0]{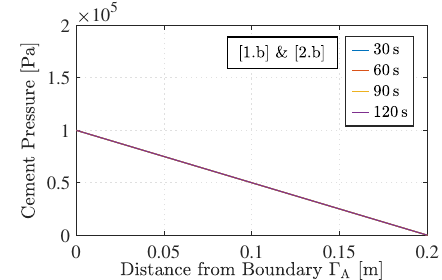}
\caption{Profiles of the cement pressure. Common results with Brooks-Corey (top) and linear relative permeability factors (bottom)}
\label{fig:results-pressure}
\end{figure}

The results of scenarios 1 and 2 differ only in terms of the temperature fields. The common results of both scenarios are depicted in Figures~\ref{fig:results-saturation}--\ref{fig:results-pressure}.

In Figure~\ref{fig:results-saturation}, profiles of the cement and marrow saturation are depicted. The saturations sum up to $1.0$ and evolve according to the employed relative permeability factors.\\ Focusing on the cement saturation, with Brooks-Corey relative permeability factors, a shock front propagates in flow direction followed by a rarefaction fan. With linear relative permeability factors, no shock front is elicited. Instead, a sigmoid-like step propagates in flow direction.

In Figure~\ref{fig:results-darcy}, profiles of the cement and marrow Darcy velocity in flow direction are depicted. The marrow Darcy velocity complements the cement Darcy velocity such that their sum is identical to $v^C$.\\
The cement Darcy velocity assumes the prescribed value $v^C$ at the inflow boundary. With Brooks-Corey relative permeability factors, the cement Darcy velocity decreases monotonically up to a shock front, which coincides with the shock front of the cement saturation. Past the shock front, the cement Darcy velocity it is zero.\\
With linear permeability factors, the Darcy velocities are directly proportional to the fluid saturations. The Darcy velocities are zero at the respective residual saturations.

In Figure~\ref{fig:results-pressure}, profiles of the cement pressure are depicted. Since we prescribe negligible capillary pressure, we omit a visualisation of the marrow pressure and pore pressure. They elicit the same behaviour as the cement pressure, as follows.\\
If Brooks-Corey relative permeability factors are employed, the pressure assumes its highest value at the inflow boundary and decreases monotonically to the value zero at the outflow boundary. Starting at the inflow boundary, the pressure decreases non-linearly up to the location of the shock front of the cement saturation. Past this point, the pressure decreases linearly. Temporally, the pressure increases at the inflow boundary as time continues, whereas the pressure past the shock front of the cement saturation remains constant.\\
If linear relative permeability factors are employed, the pressure decreases linearly along flow direction, while remaining constant in time.

\subsection{Unique Results}\label{subsec:results1}
\begin{figure}[htbp]
\includegraphics[scale=1.0]{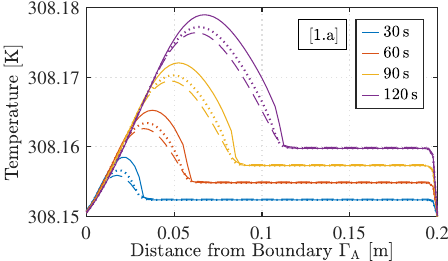}
\includegraphics[scale=1.0]{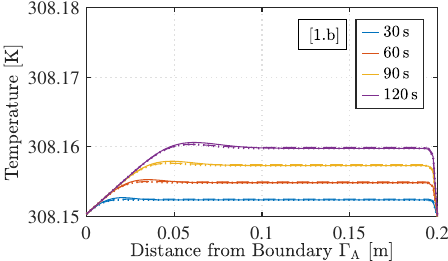}
\caption{Profiles of the cement (---), bone ($\cdots\:\!$) and marrow (-$\,$-) temperature. Results of scenario 1, with Brooks-Corey (top) and linear relative permeability factors (bottom)}
\label{fig:results-theta1} 
\end{figure}

\begin{figure}[htbp]
\includegraphics[scale=1.0]{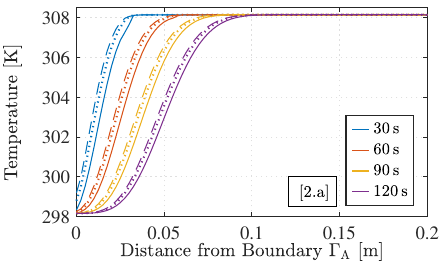}
\includegraphics[scale=1.0]{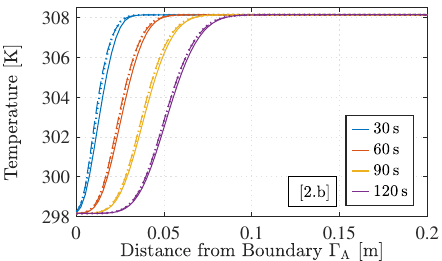}
\caption{Profiles of the cement (---), bone ($\cdots\:\!$) and marrow (-$\,$-) temperature. Results of scenario 2, with Brooks-Corey (top) and linear relative permeability factors (bottom)}
\label{fig:results-theta2}
\end{figure}

The simulation results distinguishing scenarios~1 and~2 are depicted in Figures~\ref{fig:results-theta1} and~\ref{fig:results-theta2}, respectively.

In Figure~\ref{fig:results-theta1}, profiles of the constituent temperatures are shown, resulting from the simulations of scenario 1.
As is depicted, the temperature increases inside the simulation domain.\\
If Brooks-Corey relative permeability factors are employed, starting at the inflow boundary, the temperature profiles are hill-shaped up to the location of the shock front of the corresponding cement saturation profile. Between this point and the outflow boundary the temperature profiles are constant in space. The temperature increases as time continues at both the hill-shaped and spatially constant part.\\
If linear relative permeability factors are employed, the hill-shaped parts are less pronounced. The remaining behaviour is identical.\\
In either case, all constituent temperatures elicit qualitatively identical but temporally delayed behaviour. In particular, the evolution of the marrow temperature trails behind the evolution of the solid temperature which trails behind the evolution of the cement temperature.\\
The overall temperature elevation is small. With Brooks-Corey relative permeabilities, the depicted profiles deviate from the initial temperature by $0.05\,\mathrm{K}$ at most. With linear relative permeabilities, the highest deviation reduces to $0.01\,\mathrm{K}$, approximately.

In Figure~\ref{fig:results-theta2}, profiles of the constituent temperatures are shown, resulting from the simulations of scenario 2. The temperature profiles have a step- or sigmoid-like shape and propagate in flow direction as time continues. The marrow temperature trails behind the solid temperature which trails behind the cement temperature. At the inflow boundary, the temperatures approach the temperature of the inflowing cement as time continues. Past the location of the shock front or step of the cement saturation, all temperatures assume their initial value.\\
After 120 seconds, with Brooks-Corey relative permeabilities, the constituent temperatures deviate from each other by at most $0.93\,\mathrm{K}$. With linear relative permeabilities, the highest temperature deviation after 120 seconds is determined as $0.36\,\mathrm{K}$.

\subsection{Discussion}
As per construction of the simulation scenarios, a constant cement volume influx is prescribed at the inflow boundary and the cement pressure is determined such that this volume influx is maintained. As such, we consider a rate-controlled cement injection.\\
Since we model all relevant material properties as temperature-independent, the simulation results are invariant to changes of the temperature; with exception of the temperature fields themselves.\\
The temperature fields of scenario 1 show the extent of energy dissipation, as caused by the dissipation terms $\omega^\alpha$. The elicited temperature elevations are negligible for our application.\\
The temperature fields of scenario 2 show behaviour in line with our expectations. The injection of cold cement into a warmer porous medium causes a non-uniform temperature distribution.\\
Since heat transfer occurs at a finite rate, the constituent temperatures differ. The temperature differences are negligible in our simulations, such that we may assume LTE conditions.\\
On the other hand, the injection rate we consider is rather slow in comparison to real applications~\cite{Trivedi2022}. Higher flow rates may yield non-negligible temperature differences and, by definition, will increase the extent of energy dissipation.

Leaving the behaviour of the temperature fields aside, if linear relative permeability factors are employed, our results imply that the required injection pressure for cement injection remains constant as long as the fluid viscosities are identical. In contrast, with Brooks-Corey relative permeabilities, the injection pressure increases significantly over time despite the employed fluid viscosities being identical.\\
Experimental investigations do not suggest a significant increase of the injection pressure over time~\cite{Trivedi2022}; considering time frames relevant for our application. However, our simulations do not account for effects of non-Newtonian fluid viscosities, such that a direct comparison to these experiments is not meaningful.

\section{Conclusion}
In this work, we presented a multiphase continuum mechanical model for simulating vertebroplasty. We derived our model following the principles of the TPM, assuming local thermal non-equilibrium conditions, in particular.\\
While some of our derivation steps deviate from typical approaches, our model is in agreement with other TPM-based models if restricted to local thermal equilibrium conditions.\\
We performed simulations relevant for our application. We observed no physically unreasonable behaviour in our simulation results. As such, we claim our model to be thermodynamically consistent, despite the employment of the Coleman and Noll procedure; as opposed to the rigorous but cumbersome Liu-M\"uller procedure~\cite{Ehlers02}.\\
Note, however, that we require the negligibility of capillary forces in order to make our claim. Under this condition, the filter laws of our model reduce to the extended Darcy filter law.\\
In future work, we will incorporate a consistent description of the bone cement curing into our model and we will investigate the necessity of our local thermal non-equilibrium assumption, in particular, considering the influence of the cement curing process as a heat supply.\\
If necessary, a two-scale approach as described in \cite{Ricken.2022} will also be developed and applied.

\backmatter

\bmhead{Acknowledgments}

We thank the Deutsche Forschungsgemeinschaft (DFG, German Research Foundation) for supporting this work by funding - EXC2075 - 390740016 under Germany's Excellence Strategy. In addition, we thank the DFG for supporting this work via the projects 463296570 and 504766766.  We acknowledge the support by the Stuttgart Center for Simulation Science (SimTech).

\section*{Declarations}

\begin{itemize}
\item Funding 'Funded by the Deutsche Forschungsgemeinschaft (DFG, German Research Foundation) - Project Number 327154368 - SFB 1313.'
\item Conflict of interest 'The authors have no relevant financial or non-financial interests to disclose.'
\item Ethics approval 'Not applicable'
\item Consent to participate 'Not applicable'
\item Consent for publication 'Not applicable'
\item Availability of data and materials 'In figures.'
\item Code availability 'No.'
\item Authors' contributions 'The manuscript was written by J.-S. L. V\"olter. A fundamental part of the numerical implementation was provided by Z. Trivedi. All authors have reviewed the manuscript.'
\end{itemize}

\begin{appendices}
\section{Convergence Study}\label{subsec:convergence}
We employ linear shape functions for all primary variables, as opposed to Taylor-Hood elements. Further, advective transport of scalar quantities is discretised employing upwinding. This is sufficient for our simulations to converge as we show in the following.

\begin{figure}[htbp]
\centering
\includegraphics[scale=0.311]{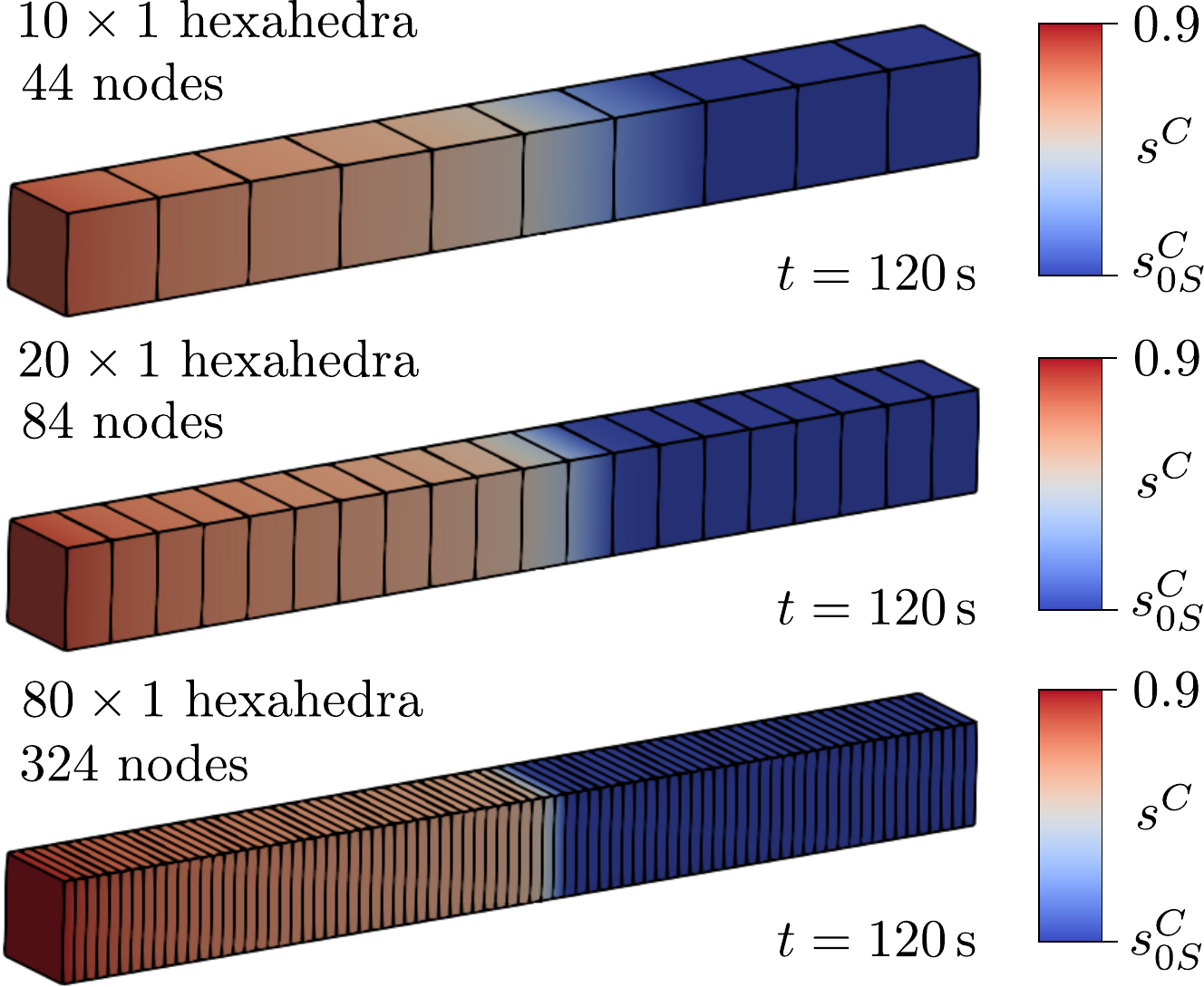}
\caption{Cement saturation after 120 seconds for selection of hexahedral finite element meshes. Results of case [2.a]}
\label{fig:convergence1}   
\end{figure}

Employing the finite element meshes depicted in Figure~\ref{fig:convergence1} as well as finer and coarser meshes, we repeatedly simulate scenario [2.a] with a fixed time-step size of $1.0$ seconds. For the finest mesh, we consider a time-step size of $0.5$ seconds.\\
We consider six rectangular hexahedral meshes of varying resolution along the flow direction. The finest mesh (not visualised) has a spatial resolution of 320 equidistant steps along flow direction.

\begin{figure}[htbp]
\includegraphics[scale=1.0]{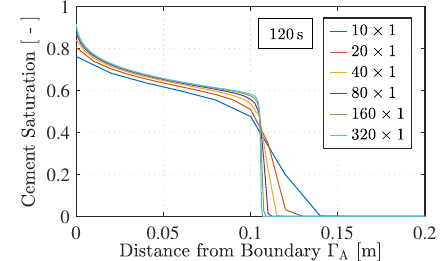}
\includegraphics[scale=1.0]{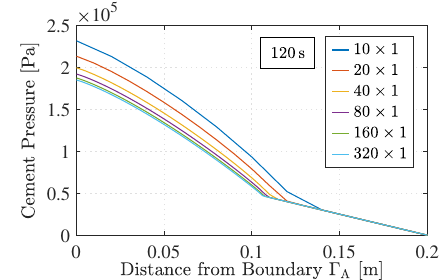}
\includegraphics[scale=1.0]{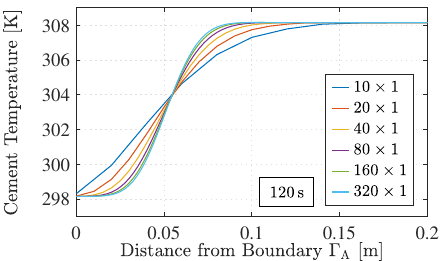}
\caption{Profiles of the cement saturation, cement pressure and cement temperature after 120 seconds. Results of case [2.a] for different hexahedral meshes}
\label{fig:convergence-saturation}    
\end{figure}

In Figure~\ref{fig:convergence-saturation}, profiles of the cement saturation are depicted after 120 seconds of simulated time for all meshes we consider. In Figure~\ref{fig:convergence1}, the corresponding scalar fields of the cement saturation are shown for selected meshes. The profiles are in agreement with the relative permeability factors, eliciting a shock front followed by a rarefaction fan. With increasing mesh-refinement along flow direction, the width of the shock front decreases and the profiles converge locally. Considering profiles of the cement pressure and temperature, similar behaviour can be observed. These profiles are visualised in Figure~\ref{fig:convergence-saturation}. We omit a description.\\
As global indicator for grid-convergence, we consider the total cement volume within the simulation domain. We predict it analytically as
\begin{equation}\label{eq:analyticvolume}
n^C_{0S}\times (2\,\mathrm{cm})^2 \times 20\,\mathrm{cm}  + v^C\times (2\,\mathrm{cm})^2\times t.
\end{equation}

\begin{figure}[htbp]
\includegraphics[scale=1.0]{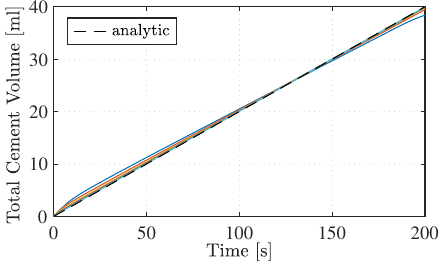}
\vspace{0.1cm}
\includegraphics[scale=1.0]{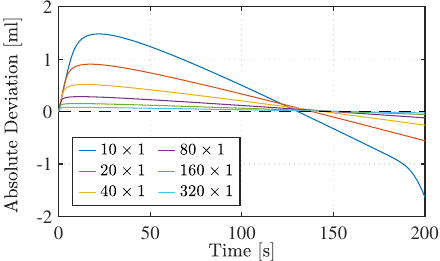}
\caption{Evolution of total cement volume and absolute deviation from analytical prediction. Results of case [2.a] employing various hexahedral meshes (---) as well as the analytic prediction  (-\,-)}
\label{fig:convergence-volume}   
\end{figure}

In Figure~\ref{fig:convergence-volume}, the temporal evolution of the total cement volume for the above simulations is visualised. As depicted, the total volume starts at the residual volume defined by the initial cement volume fraction $n^C_{0S}$ and increases monotonically as time continues. Relative to the analytic prediction~(\ref{eq:analyticvolume}), the numerical solution overestimates the cement volume initially and decreases thereafter. We attribute the decrease to numerical diffusion caused by upwinding. A sudden decrease occurs when the shock front of the cement saturation reaches the outflow boundary.\\
The numerical results approach the analytic prediction with increasing mesh-refinement along flow direction. For the finest mesh we consider ($320\times 1$ hexahedra) the deviation is below $0.08\,\mathrm{ml}$ which corresponds to $0.1\,\%$ of the total volume of the simulation domain. For our simulations we employ a coarser mesh ($80\times 1$ hexahedra), yielding a deviation below $0.28\,\mathrm{ml}$ which corresponds to $0.35\,\%$ of the total volume of the simulation domain. For the purpose of the simulations in this work, this is sufficiently accurate.

\begin{figure}[htbp]
\includegraphics[scale=1.0]{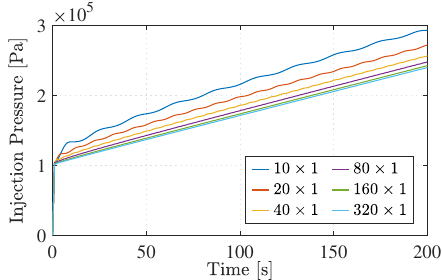}
\caption{Evolution of injection pressure. Results of case [2.a] employing various hexahedral meshes}
\label{fig:convergence-pressure}   
\end{figure}

In Figure~\ref{fig:convergence-pressure}, the evolution of the cement pressure at the inflow boundary is visualised. On average, the pressure rises monotonically but oscillates temporally. With increasing mesh-refinement along flow direction, the average increase of the pressure becomes linear in time. This is in agreement with the almost linear increase of the total cement volume and the definition of the relative permeability factors.

\end{appendices}

\bibliography{sn-bibliography}

\end{document}